\documentclass[a4paper,pre,reqno,superscriptaddress,twocolumn,
showpacs,footinbib]{revtex4-1}
\usepackage[utf8]{inputenc}
\usepackage{amsmath}
\usepackage{amsfonts}
\usepackage{amssymb}
\usepackage{makeidx}
\usepackage{graphicx}
\usepackage{color}
\usepackage{mathptmx}
\usepackage{changes,enumitem,todonotes}
\usepackage{soul}
\usepackage[mathscr]{euscript}
\usepackage{tikz}
\usepackage[normalem]{ulem}
\usetikzlibrary{patterns}
\usetikzlibrary{decorations}
\usetikzlibrary{%
  lindenmayersystems,
  decorations.pathmorphing,
  decorations.markings,
  shapes.geometric,
  calc%
}
\usetikzlibrary{snakes}

\newcommand{\ie}{{i.e.}}

\def\bea{\begin{eqnarray}}
\def\eea{\end{eqnarray}}
\def\ba{\begin{array}}
\def\ea{\end{array}}
\def\n{\nonumber}

\def\la{\langle}
\def\ra{\rangle}

\definecolor{ao(english)}{rgb}{0.0, 0.5, 0.0}
\def\o{\omega}

\usepackage[utf8]{inputenc}
\usepackage{amsmath}
\usepackage{amsfonts}
\usepackage{amssymb}

\begin{document}
\title{Stationary states of activity-driven harmonic chains}
\author{Ritwick Sarkar$^1$, Ion Santra$^2$, Urna Basu}
\affiliation{ S. N. Bose National Centre for Basic Sciences, Kolkata 700106, India}
\affiliation{ Raman Research Institute, Bengaluru 560080, India}

\begin{abstract}
We study the stationary state of a chain of harmonic oscillators driven by two active reservoirs at the two ends. These reservoirs exert correlated stochastic forces on the boundary oscillators which eventually leads to a nonequilibrium stationary state of the system. We consider three most well known dynamics for the active force, namely, active Ornstein-Uhlenbeck process, run-and-tumble process and active Brownian process, all of which have exponentially decaying two-point temporal correlations but very different higher order fluctuations. We show that irrespective of the specific dynamics of the drive, the stationary velocity fluctuations are Gaussian in nature with a kinetic temperature which remains uniform in the bulk. Moreover, we find the emergence of an `equipartition of energy' in the bulk of the system---the bulk kinetic temperature equals the bulk potential temperature in the thermodynamic limit. We also calculate the stationary distribution of the instantaneous energy current in the bulk which always shows a logarithmic divergence near the origin and asymmetric exponential tails. The signatures of specific active driving become visible in the behavior of the oscillators near the boundary. This is most prominent for the RTP and ABP driven chains where the boundary velocity distributions become non-Gaussian and the current distribution has a finite cutoff.   
\end{abstract}
\maketitle

\section{Introduction}

The study of nonequilibrium steady states (NESS) of extended systems driven by equilibrium reservoirs has been of long standing interest. Perhaps the simplest example is that of a harmonic chain connected to two thermal reservoirs at the ends, which was studied by Rieder, Lebowitz and Lieb in 1967 \cite{RLL}. It was shown that this system reaches a Gaussian NESS which carries a constant energy current, even in the limit of thermodynamically large system size. Several generalizations of this model have been studied over the past decades, ranging from inclusion of  anharmonic interaction, pinning potential and disorders, which show non-trivial stationary state behavior including anomalous transport and non-linear temperature profile~\cite{Transportbook, DharReview2008,nakazawa,RoyDhar2008,Dhar2001,FPUT,FPUT_alternatingmass,
kundu_sanjib,kannan_12}.

An important question that arises naturally is how the stationary state of an extended system is affected when it is driven by nonequilibrium reservoirs that violate the fluctuation-dissipation relation \cite{maes2013,maes2014,maes2015,vandebroek,Iacobucci2011}. Active reservoirs are a special class of nonequilibrium reservoirs that consists of self propelled particles like bacteria or Janus beads \cite{2dbacterialbath,beacterialdynamics,engein_vact_reservoir}.  
The action of active reservoirs on single probe particles has been a topic of increasing interest over the past few years, due to their unusual emergent features like negative viscosity and modification of equipartition theorem \cite{bacterialbath2011,gopal2021,kafri2021,maggi2014,maes2020,active_bath, collapse_polymer, work_fluct, dissipation_activefluid, sup_diff_colloid, santra2022}.
 Recently the effect of active reservoirs on extended systems have been studied in a simple setting similar to the model proposed by  Rieder, Lebowitz and Lieb---an ordered chain of harmonic oscillators connected to two active reservoirs which exert exponentially correlated stochastic forces on the boundary oscillators~\cite{activity_driven_chain}. It was shown that this simple system exhibits some remarkable features like negative differential conductivity and current reversal. Both the average energy current and kinetic temperature profile, which were computed  exactly, depend only on the autocorrelation of the active force and holds true irrespective of the specific dynamics. However, the signatures of the specific dynamics of the active forces are expected to be present in the higher order fluctuations of these observables.

In this paper we study the NESS of a harmonic chain driven by different kinds of exponentially correlated active forces.
In particular, we consider three most well known active processes, namely, Active Ornstein-Uhlenbeck Process (AOUP) \cite{AOUP}, Run-and-Tumble Process (RTP) \cite{rtp1, rtp_roman} and Active Brownian Process (ABP) \cite{Howse2007,abp_fodor} to model the dynamics of the active forces. To characterize the NESS, we focus on the behavior of the energy current, velocity  and potential energy fluctuations of the oscillators.
Surprisingly, we find that the bulk properties in the NESS are universal and do not depend on the specific dynamics of the active forces. More specifically, we show that, in all the three cases, the instantaneous current distribution at the bulk has logarithmic divergence near the origin as well as asymmetric exponential tails. We also find that the velocity fluctuations of the bulk oscillators are Gaussian, which is accompanied by an `equipartition of energy' --- in the thermodynamic limit, the kinetic and potential temperatures become equal in the bulk, which we show analytically.

The signatures of the specific dynamics of the active force become visible in the behavior of the oscillators near the boundaries. In particular, we show that the velocity distributions of the boundary oscillators show different non-Gaussian features for the ABP and RTP driven chains. On the other hand, the Gaussian nature of the AOUP active force ensures that the boundary velocity fluctuations remain Gaussian in this case. The instantaneous current distributions at the boundaries show more surprising features--- for ABP and RTP drives, the boundary current distributions have semi-finite supports, which can be understood from the bounded nature of the driving forces in these cases. For AOUP, on the other hand, the boundary current distribution has exponential tails which we compute exactly.   	
	
The paper is organized as follows. In the next section we introduce the setup and give a brief summary of our results. Sections~\ref{sec:temp}  and \ref{sec:veldist}
are devoted to the study  of the temperature profile and velocity distributions of the oscillators. The behavior of the current distributions  is discussed in Sec.~\ref{sec:cur}. We conclude with some general remarks in Sec.~\ref{sec:concl}. 

\section{Model and Results}\label{sec:model}
\begin{figure}[t]
\includegraphics[width=9cm]{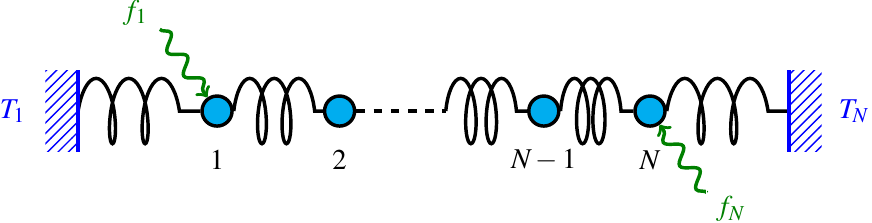}
\caption{Schematic representation of a harmonic chain of oscillators connected to two nonequilibrium reservoirs at the two ends. Apart from the usual thermal noise the boundary oscillators are driven by auto-correlated active forces $f_1(t)$ and $f_N(t)$.}
\label{fig:model}
\end{figure}

We consider a chain of $N$ oscillators, each with mass $m$, connected by springs of stiffness $k$. The chain is connected to two active reservoirs which exert exponentially correlated stochastic forces on the boundary oscillators in addition to the usual white noise and dissipative forces coming from thermal reservoirs [see Fig.~\ref{fig:model}]. The displacement $x_l$ of the $l$-th oscillator from its equilibrium position follows the equations of motion,
\begin{subequations}
\bea
m\dot{v}_1 &=&-k(2x_1-x_2)-\gamma\,\dot{x_1}+\xi_1(t)+f_1(t),\label{eom1} \\
m\dot{v}_l&=& -k(2x_l-x_{l-1}-x_{l+1}),~~~\forall\, l\in[2,N-1],\label{eom2} \\
m\dot{v}_N&=&-k(2x_N-x_{N-1})-\gamma\,\dot{x_N}+\xi_N(t)+f_N(t),~~~\label{eom3}
\eea\label{ts}
\end{subequations}
where $v_l = \dot x_l$ and we have assumed fixed boundary condition, $x_0=x_{N+1}=0$. The white noises $\xi_1$ and $\xi_N$  acting on the boundary oscillators denote the forces from the thermal components of the reservoirs which satisfy the fluctuation-dissipation relation \cite{Kubo},
\bea
\la \xi_i(t) \xi_j(t') \ra=2 \gamma T_j\, \delta_{i j}\, \delta(t-t').
\eea
Here $T_1$ and $T_N$ denote the temperatures of the reservoirs and for simplicity we have assumed that the dissipation coefficient  $\gamma$ is the same for both the reservoirs. The active  forces $f_j(t)$ are assumed to be exponentially correlated colored noises,
\bea
\la f_i(t)f_j(t')\ra =\delta_{ij} \, a_j^2 \exp(-|t-t'|/\tau_j), \label{eq:fcorr}
\eea
where $\tau_{1,N}$ measure the activity of the reservoirs.

The linear Langevin equations~\eqref{ts} can be straightforwardly solved in the frequency domain to obtain \cite{Dhar2001}, 
\bea
x_l(t)=\int_{-\infty}^{\infty} \frac{d \omega}{2 \pi} e^{-i \omega t} \left[ G_{l1}(\omega)\tilde{f}_1(\omega) + G_{lN}(\omega)\tilde{f}_N(\omega)\right],~~~\label{eq:xl}
\eea
where $\tilde{f}_j(\omega)$ is the Fourier transform of $f_j(t)$ with respect to $t$ and $G(\omega)$ is the Green's function matrix; see Appendix~\ref{calc_appendix} for the detailed solution. 
%
%
%
%
%
%
%
%

 Clearly, the stationary state distribution of $\{x_l, v_l\}$ would depend on the statistical properties of the active force $f_j(t)$ through $\tilde{f}_j(\omega)$. From Eq.~\eqref{eq:xl}, it is clear that the two-point dynamical correlations of physical observables which are linear in $x_l(t)$, involve only the two-point correlation $\la \tilde{f}_i(\omega)\tilde{f}_j(\omega')\ra=\delta_{ij} \tilde{g}(\omega,\tau_j)\delta(\omega+\omega')$, where $\tilde{g}(\omega,\tau_j)$ is the frequency spectrum of the active force and is given by a Lorentzian,
\bea
\tilde{g}(\omega,\tau_j) = \frac{2 a_j^2 \tau_j }{1+\omega^2 \tau_j^2}.
\label{eq: lorentzian}
\eea
In the following we consider three different dynamical processes which correspond to very different fluctuations of $f_j(t)$, although each has an exponentially decaying autocorrelation of the form Eq.~\eqref{eq:fcorr}.

\begin{itemize}[leftmargin=*] 

\item [I.]{\bf Active Ornstein-Uhlenbeck Process (AOUP)}: We first consider the scenario where the active force at each boundary undergoes an independent  Ornstein-Uhlenbeck process \cite{AOUP, aoup},
\bea
\dot f_j(t) = -\frac 1 {\tau_j} f_j + \sqrt{\frac{2D_j}{\tau_j^2}}\, \eta_j(t), \label{eq:aoup}
\eea
where $\eta_j(t)$ is a Gaussian white noise with $\la \eta_j(t) \ra =0$ and  $\la \eta_j(t) \eta_j(t') \ra =\delta(t-t')$; the diffusion constant $D_j$ denotes the strength of the noise.  The linear nature of the process and the Gaussian nature of the noise leads to a Gaussian propagator for the active force $f_j(t)$,
{\bea
{\cal P}(f_j,t|f_j',t') = \frac{\exp{\left( -\frac{\tau_j}{2 D_j}\frac{(f_j-f'_j e^{-(t-t')/\tau_j})^2}{1-e^{-2(t-t')/\tau_j}} \right)}}{\Big[2 \pi D_j(1-e^{-2(t-t')/\tau_j})\Big]^{1/2}}.\label{aoup_prop}
\eea}
Evidently, the stationary distribution of $f_j$ is also Gaussian with $\la f_j \ra=0$ and $\la f_j^2 \ra=D_j/\tau_j$. Equation~ \eqref{aoup_prop} implies that the stationary two-point correlation of the active force $\la f_j(t) f_j(t') \ra$ is given by Eq.~\eqref{eq:fcorr} with,
\bea
a_j = \sqrt{D_j/ \tau_j}. \label{eq:aj_aoup}
\eea
The linear nature of the system and the Gaussian nature of the active force $f_j$ ensures that, for the AOUP drive, the joint probability distribution of $\{ x_l,v_l \}$ is also Gaussian,
\bea
\mathscr{P}(\{ x_l,v_l \}) = \frac 1{\sqrt{(2\pi)^{2N}\det(\Sigma)}} \exp{\left[-\frac 12 W^T \Sigma^{-1} W\right ]} \label{eq:Pxv_N}
\eea
where $W^{T} = (v_1 ,\cdots, v_N, x_1, \cdots,  x_N)$ and $\Sigma$ is the corresponding $2N \times 2N$ dimensional positive-definite correlation matrix.

\begin{figure*}[th!]
\includegraphics[width=17cm]{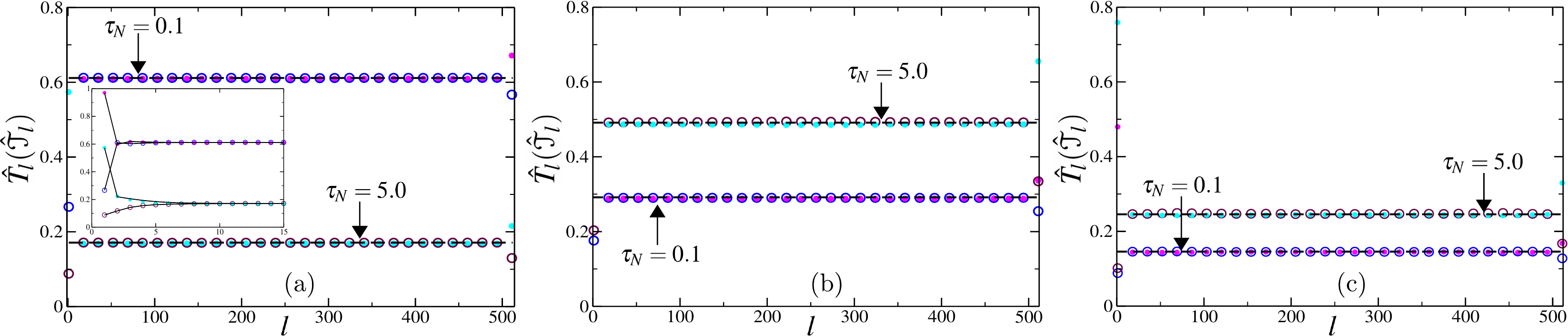}
\caption{Plot of kinetic temperature profile $\hat{T}_l$ (open symbols) and potential temperature profile $\hat{\mathscr{T}}_l$ (filled symbols) for the AOUP (a), RTP (b) and ABP (c) driven chains, for fixed $\tau_1=2.0$ and two different values of $\tau_N$. The symbols correspond to data obtained from numerical simulations with a chain of $N=512$ oscillators and, $D_1=D_N=1$ in (a), $A_1=A_N=1$ in (b) and $D_1^R=D_N^R=1$ in (c). The other parameters are $\gamma=1=k=m$.  The black dashed lines correspond to the value of the bulk temperature according to Eq.~\eqref{bulk_temp}. The inset in (a) shows zoomed in temperature profiles near the left boundary. The solid black lines in the inset corresponds to the analytical predictions of  $\hat {\mathscr T}_l$ [see Appendix \ref{appendix_pot}] and $\hat {T}_l$ \cite{activity_driven_chain}.}
\label{fig:energy_prof}
\end{figure*}

\item[II.] {\bf Run-and-tumble process (RTP)}: In this case we consider the active force $f_j(t)$ to be a dichotomous noise similar to the famous Run-and-Tumble process \cite{rtp1,rtp},
\bea
f_j(t) = A_j \sigma_j(t),
\eea
where $\sigma_j(t)$ alternates between $1$ and $-1$ with rate $\alpha_j$. In this case $f_j=\pm A_j$ can take only two discrete values and the corresponding propagator is given by \cite{drabp},
\bea
P(f_j,t|f_j',t')=\frac{1}{2}\left( 1+f_j f_j' e^{-2 \alpha_j|t-t'|}\right).\label{rtp_prop}
\eea  
Clearly, in the stationary state, the two values of $f_j$ occur with equal probability $1/2$. It is straightforward to see that this process leads to the two point auto-correlation of the form Eq.~\eqref{eq:fcorr} with 
\bea
\tau_j  = 1/(2 \alpha_j),\quad\text{and}\quad a_j = A_j. \label{eq:aj_rtp}
\eea
However, the higher order correlation of $f_j$, computed from Eq.~\eqref{rtp_prop}, are quite different from that of the AOUP, and, in general, the stationary state distribution $\mathscr{P}(\{ x_l,v_l \})$ is expected to be non-Gaussian.

\item[III.] {\bf Active Brownian process (ABP)}: The third case refers to the scenario where the active force evolves according to the active Brownian dynamics \cite{abp_fodor,abp},
\bea
f_j(t) = A_j \cos \theta_j(t),\quad\text{with}\quad \dot \theta_j(t) = \sqrt{2 D^R_j}\, \zeta_j(t),
\eea

where $\zeta_j$ refers to a Gaussian white noise with $\la \zeta_j(t) \ra =0$ and  $\la \zeta_j(t) \zeta_j(t') \ra =\delta(t-t')$. Clearly, $\theta_j(t)$ undergoes a standard Brownian motion which leads to a Gaussian propagator \cite{drabp},
\bea
P(\theta_j,t|\theta_j',t)=\frac{1}{\sqrt{4 \pi D_j^R |t-t'|}}\exp{\left(-\frac{(\theta_j-\theta_j')^2}{4 D^R_j|t-t'|}\right)}.
\eea
Corresponding distribution for $f_j=A_j \cos{\theta_j(t)}$ eventually reaches a stationary state, 
\bea
{\cal P}_{\text{st}}(f_j)=\frac{1}{\pi\sqrt{A_j^2-f_j^2}}.
\eea
The auto-correlation $\la f_j(t) f_j(t') \ra$ is given by Eq.~\eqref{eq:fcorr} with 
\bea
\tau_j=1/D_j^R,\quad\text{ and } a_j=A_j/\sqrt{2}.\label{eq:aj_abp}
\eea
However, the higher order correlation for $f_j$ for this case is different than that of both AOUP and RTP and the stationary state weight $\mathscr{P}(\{x_l,v_l\})$  is expected to be non-Gaussian as well as different from that in the RTP driven case.  
\end{itemize}
 
Clearly, despite having the same two-point auto-correlation given by Eq.~\eqref{eq:fcorr}, the dynamical nature of the active force $f_j$ is very different for all the three cases. We expect to see the signatures of these specific dynamics in the  stationary state of the different activity driven harmonic chains.

To characterize the stationary state properties of the activity driven chain we focus on the potential energy, local velocity, and current fluctuations in the harmonic chain, both in the bulk and at the boundaries. We support our analytical results with the help of numerical simulation using stochastic second order Runge-Kutta  algorithm \cite{rk2_wht,lang_int}. Note that, for a harmonic chain, energy current in the stationary state splits into two components--- a thermal one $J_\text{therm}$, proportional to the temperature difference $(T_1-T_N)$ of the thermal reservoirs, and an active one  $J_{\text{act}}$, which depends on the activity driving~\cite{activity_driven_chain}. Since we are mainly interested in characterizing the activity driven stationary state, we use $T_1=T_N=0$ for the remainder of the paper.  Before going into the details of the computation, we first present a brief summary of our main results.\\
 
\noindent {\bf Temperature profile:} We first compute the local potential temperature profile, defined as,
\bea
\hat {\mathscr T}_l = 2 U_l, \label{eq:TU_def}
\eea
where $U_l$ denotes the average potential energy of the $l$-th oscillator. We show that, $\hat {\mathscr T}_l$ becomes uniform in the bulk (i.e., for $1 \ll l \ll N$) in the thermodynamic limit $N\to \infty$ and the bulk potential temperature value, given by,
\begin{eqnarray}
\hat{\mathscr{T}}_{\text{bulk}}= \frac{a_1^2 \tau_1}{2\gamma \sqrt{1+\frac{4 \tau_1^2 k}{m}}}+\frac{a_N^2 \tau_N}{2\gamma \sqrt{1+\frac{4 \tau_N^2 k}{m}}},\label{bulk_temp}
\end{eqnarray}
which is the same as the bulk kinetic temperature $\hat{T}_\text{bulk}$ computed earlier \cite{activity_driven_chain}, which indicates the existence of an `equipartition of energy'.\\

\noindent {\bf Velocity distribution:} We also measure the stationary probability distribution $P(v_l)$ of the velocities of the oscillators and show that, surprisingly, in the limit of thermodynamic size, for any activity of the reservoirs, the velocity distributions of the bulk oscillators are Gaussian with width $\hat{T}_\text{bulk}$, irrespective of the dynamics of the active force. The velocity distributions of the oscillators near the boundaries, however, are non-Gaussian for ABP and RTP driven chains, and depend on the specific driving dynamics.\\

\noindent {\bf Current distribution:} Another observable of immense importance is the energy current flowing through the system. We show that, for the bulk oscillators, $P(\mathscr{J}_l)$, the probability distribution of the instantaneous current $\mathscr{J}_l$, flowing from the $(l-1)$-th to the $l$-th oscillator, exhibits certain universal features, irrespective of the specific dynamics of the active force: The distribution diverges logarithmically for $|\mathscr{J}_l| \to 0$ and  shows asymmetric exponential decay for large $\mathscr{J}_l$,
\bea
P(\mathscr{J}_l)\simeq \left\{ 
\begin{split}
&-\frac{\ln{|\mathscr{J}_l|}}{\sqrt{\pi^2 g_l}}\quad\text{for } |\mathscr{J}_l|\rightarrow 0,
\cr
&\frac{1}{\sqrt{2 \pi u_l |\mathscr{J}_l|}}\exp{\left[\frac{J_\text{act}\,\mathscr{J}_l-u_l|\mathscr{J}_l|}{g_l}\right]}\quad\text{for } |\mathscr{J}_l|\gg 1,
\end{split}
\right.
\eea
where $J_{\text{act}}$ and $g_l$ and $u_l$ are defined in Eqs.~\eqref{eq:J} and \eqref{eq:glul_prim}.

In fact, the Gaussian nature of the stationary state of the AOUP driven chain allows us to exactly compute the stationary current distribution in the bulk,
\begin{equation}
P(\mathscr{J}_l) = \frac{1}{\sqrt{\pi^2 g_{l}}}e^{\frac{J_\text{act}}{g_{l}} \mathscr{J}_l} K_0\left(\frac{u_{l}}{g_{l}}|\mathscr{J}_l|\right),
\end{equation}
where $K_0(z)$ is the zeroth order modified Bessel function of the  second kind~\cite{DLMF}.

We also compute the boundary current distribution for the AOUP driven chain which has the same qualitative shape as the bulk current distribution. For RTP and ABP driven chains, however, the boundary current distributions are strikingly different, which we measure numerically.

\section{Temperature profile}\label{sec:temp}

It is often convenient to consider a local `kinetic temperature' for  driven oscillator chains, which can be defined as the average kinetic energy of the $l$-th oscillator,
\bea
\hat{T}_l=m \la \dot{x}_l^2 \ra. \label{eq:Tkin_def}
\eea
For an activity driven harmonic chain, it has been shown that
the kinetic temperature attains a uniform value 
\bea
\hat T_\text{bulk} = 
\frac{a_1^2 \, \tau_1}{2\gamma \sqrt{1+\frac{4 \tau_1^2 k}{m}}}+\frac{a_N^2 \, \tau_N}{2\gamma \sqrt{1+\frac{4 \tau_N^2 k}{m}}},\label{eq:bulk_kin_temp}
\eea
in the bulk, with an exponentially decaying boundary layer \cite{activity_driven_chain}. 

For a harmonic chain, one can also define a local `potential temperature',  $\mathscr {\hat T}_l$, from the average potential energy of the $l$-th oscillator $U_l$ [see Eq.~\eqref{eq:TU_def}], defined as,
\bea
U_l=
\left \{
\begin{split}
      \frac{k}{4} \Big[2\la x_{l}^2(t)\ra +\la (x_{l+1}(t)-x_l(t))^2\ra \Big] \quad \text{for } l=1\\
      \frac{k}{4} \Big[ \la (x_{l-1}(t)-x_l(t))^2\ra +\la (x_l(t)-x_{l+1}(t))^2\ra \Big] \cr \forall l\in [2,N-1]  \\
      \frac{k}{4} \Big[2\la x_{l}^2(t)\ra +\la (x_{l}(t)-x_{l-1}(t))^2\ra \Big] \quad \text{for } l=N.
      \end{split} 
\right. ~~    
&&       \label{eq:Uav_def}
\eea  
To compute $U_l$, we need position correlations of the form $\la x_{l}(t) x_{n}(t) \ra$ in the stationary state, for $n=l,l\pm1$. From Eq.~\eqref{eq:xl} we have,
\begin{align}
\la x_{l}(t) x_{n}(t) \ra = \int_{-\infty}^{\infty}\frac{d \omega}{2 \pi}\, \Big[G_{l1} G_{1n}^* \,\tilde{g}(\omega,\tau_1)+G_{lN} G_{Nn}^* \, \tilde{g}(\omega,\tau_N) \Big],
\label{crosscorr_x}
\end{align}
where, $\tilde{g}(\omega,\tau_j)$, given in Eq.~\eqref{eq: lorentzian}, denotes the Lorentzian spectrum of the active force.

These correlations can be computed exactly using the explicit form of $G_{ln}(\omega)$. The details of the computation are provided in Appendix~\ref{appendix_pot}; here we quote the main results. It turns out, that the average potential energy can be expressed as the sum of two contributions from the two reservoirs,
\bea
{U}_l=\frac{k}{4}[ \mathscr{U}_1(l,\tau_1)+ \mathscr{U}_N(l,\tau_N)]. \label{eq:bulk_pot_temp_U}
\eea
Here $\mathscr{U}_1(l,\tau_1)$ and $\mathscr{U}_N(l,\tau_N)$ are the contributions form left and right reservoirs respectively [see  Appendix~\ref{appendix_pot} ]. We find that, $ \mathscr{U}_j(l,\tau_j) $ for bulk oscillators, is independent of $l$ in the thermodynamic limit $N \to \infty$, 
\bea
\mathscr{U}_j(l,\tau_j) &=& \frac{m\tau_j a^2_j}{k \gamma \pi}\int^{\pi}_0  \frac{dq}{m+2 k \tau_j^2(1-\cos{q})}\cr 
&=&\frac{a_j^2\tau_j}{k\gamma\sqrt{1+\frac{4 k \tau_j^2}{m}}},
\eea
where $j=1,N$. Consequently, the potential temperature profile $\mathscr{\hat T}_l$ attains a uniform value $\hat{\mathscr{T}}_\text{bulk}$ in the bulk. In fact, from Eq.~\eqref{eq:bulk_pot_temp_U}, \eqref{eq:bulk_kin_temp} and the above equation, it is clear that 
\bea
\hat{\mathscr{T}}_\text{bulk} = \hat{T}_\text{bulk}, \label{kin_temp}
\eea
i.e., the bulk kinetic and potential temperatures are identical in the thermodynamic limit. Note that, Eq.~\eqref{kin_temp} holds irrespective of the specific form of the dynamics. Fig.~\ref{fig:energy_prof} shows plots of $\hat T_l$ and $\mathscr{\hat T}_l$ for AOUP, RTP and ABP for two sets of $\tau_1$ and $\tau_N$ and validates our prediction Eq.~\eqref{kin_temp}.

\begin{figure}[tb]
\includegraphics[width=8.8cm]{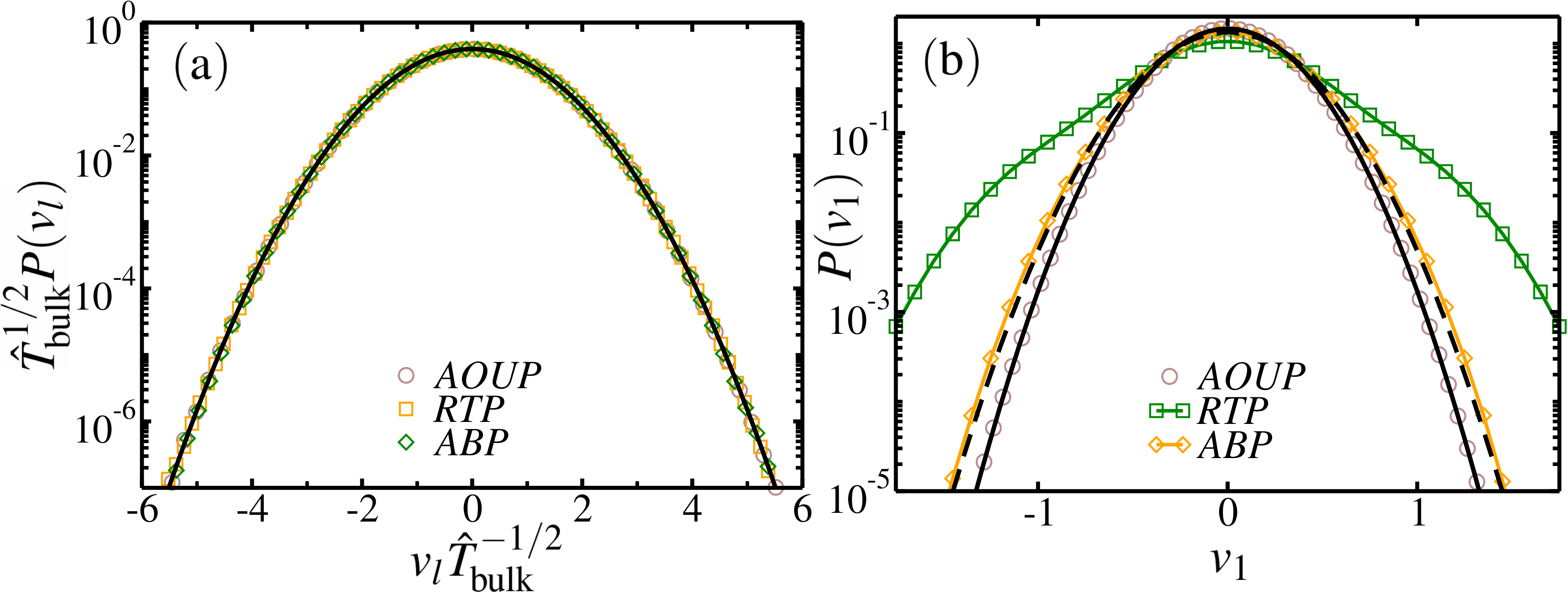}
\caption{(a) Plot of the scaled velocity distribution of the middle oscillator $N/2$ for $\tau_1=5.0$ and $\tau_N=1.50$. The data corresponding to AOUP, RTP and ABP driven chains show a perfect collapse with the standard normal distribution indicated by the solid black line. (b) Plot of the velocity distributions of the oscillators at the left boundary for the three models AOUP, RTP, and ABP, with $\tau_1=5.0$ and $\tau_N=1.50$. The black solid line corresponds to the Gaussian distribution for AOUP, and dashed line corresponds to a Gaussian with the variance $\hat{T}_1$ for ABP. For both (a) and (b) $N=512$ and $m=\gamma=k=1$. The other parameters are $D_1=D_N=1$ for AOUP,
$A_1=A_N=1$ for RTP and $D_1^R=D_N^R=1$ for ABP.}
\label{fig:v_dist}
\end{figure}

\begin{figure*}[t]
\includegraphics[width=17cm]{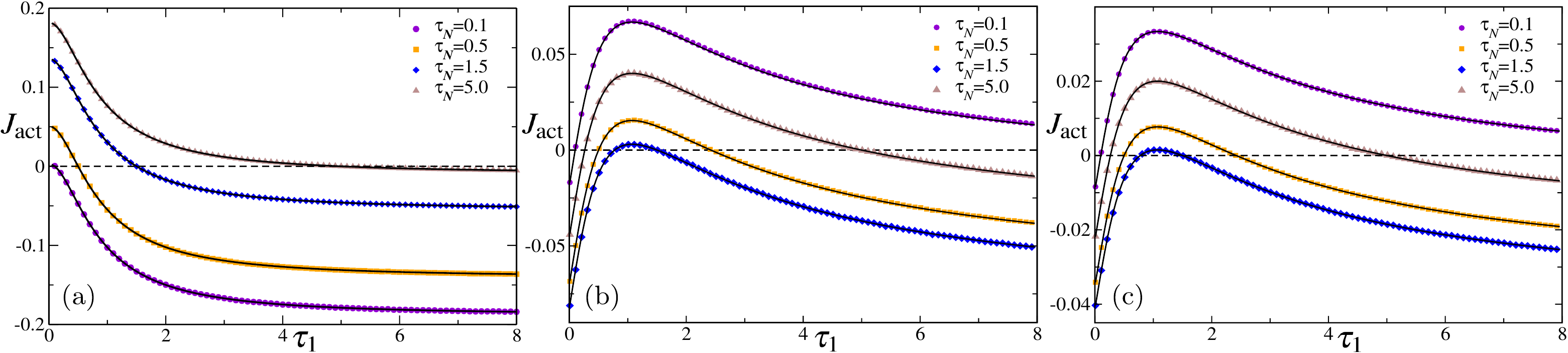}
\caption{Plot of the average  active current $J_\text{act}$ as functions of activity $\tau_1$, for AOUP (a), RTP (b) and ABP (c) driven chains for different values of $\tau_N$. Symbols corresponds to the data obtained from numerical simulations with $N=64$ oscillators and the other parameters are $D_1=D_N=1$ for AOUP,   $A_1=A_N=1$ for RTP,  $D_1^R=D_N^R=1$ for ABP and $\gamma=1=k=m$. Black solid lines correspond to Eq.~\eqref{eq:J}. }
\label{fig:current}
\end{figure*}

The potential temperatures of the oscillators near the boundaries, calculated explicitly in Appendix~\ref{appendix_pot}, are different from their respective kinetic temperatures. The difference is illustrated in the inset of Fig.~\ref{fig:energy_prof} (a) for two sets of $\tau_1$ and $\tau_N$.

\section{Velocity distributions}
\label{sec:veldist}

The probability distribution of the velocities plays an important role in the characterization of the NESS of the oscillator chain. In the presence of a thermal gradient such a system usually reaches a stationary state, where the velocity fluctuation of the $l$-th oscillator are typically Gaussian with the width given by its local kinetic temperature \cite{Transportbook,DharReview2008}. In this section we explore the fluctuation of the velocities of the individual oscillators in the presence of the different active drivings.

For the AOUP driven chain, as mentioned before, the joint probability distribution $\mathscr{P}(\{x_l,v_l\})$ is a multivariate Gaussian [see Eq.~\eqref{eq:Pxv_N}]. Consequently, the marginal velocity distribution $P(v_l)$ must also be a Gaussian,
\bea
P(v_l)=\frac{1}{\sqrt{2 \pi \hat{T}_l/m}} \exp\left (-\frac{m v_l^2}{2 \hat{T}_l}\right),~~\label{eq:gauss_vl}
\eea
for $l=1,2,\cdots N$, where $\hat{T}_l=m \la v_l^2 \ra$ is the average kinetic temperature of the $l$-th oscillator. This is illustrated in Fig.~\ref{fig:v_dist}(a) where the numerically measured velocity distribution of the middle oscillator ($l=N/2$) is plotted along with the corresponding Gaussian which shows perfect agreement.

For RTP and ABP driven chains, on the other hand, Eq.~\eqref{eq:Pxv_N} is not expected to hold. Surprisingly, however, numerical simulations show that for oscillators in the bulk, the typical velocity fluctuations are still Gaussian. This is shown in Fig. \ref{fig:v_dist} (a) where the scaled velocity distributions of the $l=N/2$-th oscillator of the RTP and ABP driven chains are compared with Eq.~\eqref{eq:gauss_vl} showing an excellent agreement. Nevertheless, the signatures of the underlying non-Gaussian stationary states become apparent in the velocity fluctuations of the oscillators near the boundaries. Figure \ref{fig:v_dist}~(b) shows a plot of the marginal distribution $P(v_1)$ of the left boundary oscillator---the obvious non-Gaussian nature of the distribution is very clear for RTP, while for ABP, the deviation from Gaussian form Eq.~\eqref{eq:gauss_vl} becomes prominent at the tails. For AOUP driven chain the boundary velocity fluctuations are also Gaussian, as expected.

\section{Current fluctuations}\label{sec:cur}

The NESS of an activity driven harmonic chain is characterized by the existence of an average energy current flowing through the system, which can be computed exactly \cite{activity_driven_chain}. Instantaneous current at the left and right boundaries $\mathscr{J}_{1}$ and $\mathscr{J}_{N+1}$ are defined as the rate of work done by left reservoir and right reservoir on the system, respectively,
\begin{eqnarray}
 \mathscr{J}_{1}=(-\gamma v_{1} + f_{1})v_{1}\text{ and, }\mathscr{J}_{N+1}=(-\gamma v_{N} + f_{N})v_{N}\label{boundary_current}.
\end{eqnarray}
The instantaneous energy current flowing from the $(l-1)$-th to $l$-th oscillator is given by,
\begin{eqnarray}
 \mathscr{J}_{l}&=&\frac{k}{2}\left(v_{l-1}+v_{l}\right) \left(x_{l-1}-x_{l} \right).\label{bulk_current}
\end{eqnarray} 
The Hamiltonian nature of the bulk dynamics ensures that in the stationary state,
\bea
\la \mathscr{J}_{1}\ra = \la \mathscr{J}_{2} \ra =\cdots =\la \mathscr{J}_{l} \ra = \cdots = -\la \mathscr{J}_{N+1} \ra=J_\text{act},
\eea where, $J_\text{act}$ is the average energy current flowing through the system. It has been shown~\cite{activity_driven_chain} that the average active current is given by a Landauer-like formula,
\bea
J_\text{act}=\gamma \int_{-\infty}^{\infty} \frac{d \omega}{2 \pi} \omega^2 |G_{1N}(\omega)|^2\Big[ \tilde{g}(\omega,\tau_1)-\tilde{g}(\omega,\tau_N)\Big],\label{eq:current_sep}
\eea
where $|G_{1N}(\omega)|^2$ denotes the phonon transmission coefficient and $\tilde{g}(\omega,\tau_j)$ corresponds to the Lorentzian spectra of the $j$-th active reservoir. The presence of the non-trivial reservoir spectra makes the activity driven current different than the thermally driven scenario, where the average current is given by,
\bea
J_\text{therm}=\big( T_1-T_N\big) \int_{-\infty}^{\infty} \frac{d \omega}{2 \pi} \omega^2 |G_{1N}(\omega)|^2.
\eea
Here $T_1$ and $T_N$ denote the temperatures of the thermal reservoirs attached at the two ends of the chain.

For a thermodynamically large chain of oscillators driven by active forces satisfying Eq.~\eqref{eq:fcorr}, the average active current is given by,
\begin{eqnarray}
J_\text{act}&=&\frac{m}{2 \gamma ^2} \left[a_1^2{\cal E}(\tau_1)-a^2_{N}{\cal E}(\tau_N)\right],~\text{with,} \label{eq:J} \n \\
{\cal E}(\tau_j)&=&\frac{\tau_j^2 k^2 \left( \sqrt{1+\frac{4\gamma^2}{mk} }-1 \right)+\gamma^2\left( 1-\sqrt{1+\frac{4 k \tau_j^2}{m} } \right)}{2 \tau_j(\tau_j ^2 k^2-\gamma^2)}.\qquad 
\eea
Note that, ${\cal E}(\tau_j) $ is nonmonotonic in $\tau_j$, and its form does not depend on the specific active force dynamics. However, $J_\text{act}$ also depends on $a_j(\tau_j)$ which makes its $\tau_j$ dependence different for the different models. In particular, for AOUP, $a_j\propto1/\sqrt{\tau_j}$, which results in an active current which monotonically decreases as function of $\tau_j$, which is illustrated in Fig.~\ref{fig:current}(a). On the other hand, for RTP and ABP, $a_j$ does not depend on $\tau_j$ resulting in a nonmonotonic behavior of $J_\text{act}$ indicating the emergence of a negative differential conductivity. This is shown in Fig.~\ref{fig:current}(b) and (c) for RTP and ABP, respectively.

More apparent signatures of the specific active force are expected to be encoded in the higher order fluctuations of the instantaneous current which we investigate next.

\subsection{Stationary distribution of $\mathscr{J}_l$ in the bulk }
\label{sec:bulk_dist}
We start with the stationary distribution of the instantaneous current $P(\mathscr{J}_l)$ for the bulk oscillators. From Eq.~\eqref{bulk_current} we can write 
\begin{eqnarray}
P(\mathscr{J}_l)=\int dv_{l-1} dv_{l} dx_{l-1} dx_{l}\, \mathscr{P}(v_{l-1}, v_{l}, x_{l-1}, x_{l})\n \\ \qquad \times  \delta\left(\mathscr{J}_l-\frac{k}{2} ( v_{l-1}+v_{l})( x_{l-1}-x_{l}) \right).
\eea
First we consider the AOUP driven chain. The Gaussian nature of the stationary state [see Eq.~\eqref{eq:Pxv_N}] in this case implies that the joint distribution of $\{v_{l-1},v_{l},x_{l-1},x_{l}\}$ is also a multivariate Gaussian,
 \begin{eqnarray}
\mathscr{P}(v_{l-1},v_{l},x_{l-1},x_{l})&=&\frac{\exp\left[ -\frac{1}{2}W_l^T\Sigma^{-1}_l W_l  \right]}{\sqrt{(2 \pi)^4 \text{det}(\Sigma_{l})}}, \label{eq: joint_dist_bulk}
\eea
where, $W_l^T=(v_{l-1}~v_{l}~x_{l-1}~x_{l})$ and $\Sigma_{l}$ is the corresponding $4\times 4$ correlation matrix [see Eq.~\eqref{eq:corrmat_bulk} in Appendix~\ref{appendix_curdist}]. To compute $ P(\mathscr{J}_l)$, it is most convenient to consider its Fourier transform with respect to $\mathscr{J}_l$ which is the moment generating function,
\bea
\langle e^{i \mu \mathscr{J}_l}\rangle = \int dv_{l-1} dv_{l} dx_{l-1} dx_{l} \,  e^{i \mu \frac{k}{2} ( v_{l-1}+v_{l})( x_{l-1}-x_{l})}  \n \\ \qquad \times \mathscr{P}(v_{l-1}, v_{l}, x_{l-1}, x_{l}). \label{eq: moment_gen_bulk}
\eea

\begin{figure}[t]
\includegraphics[width=8.8cm]{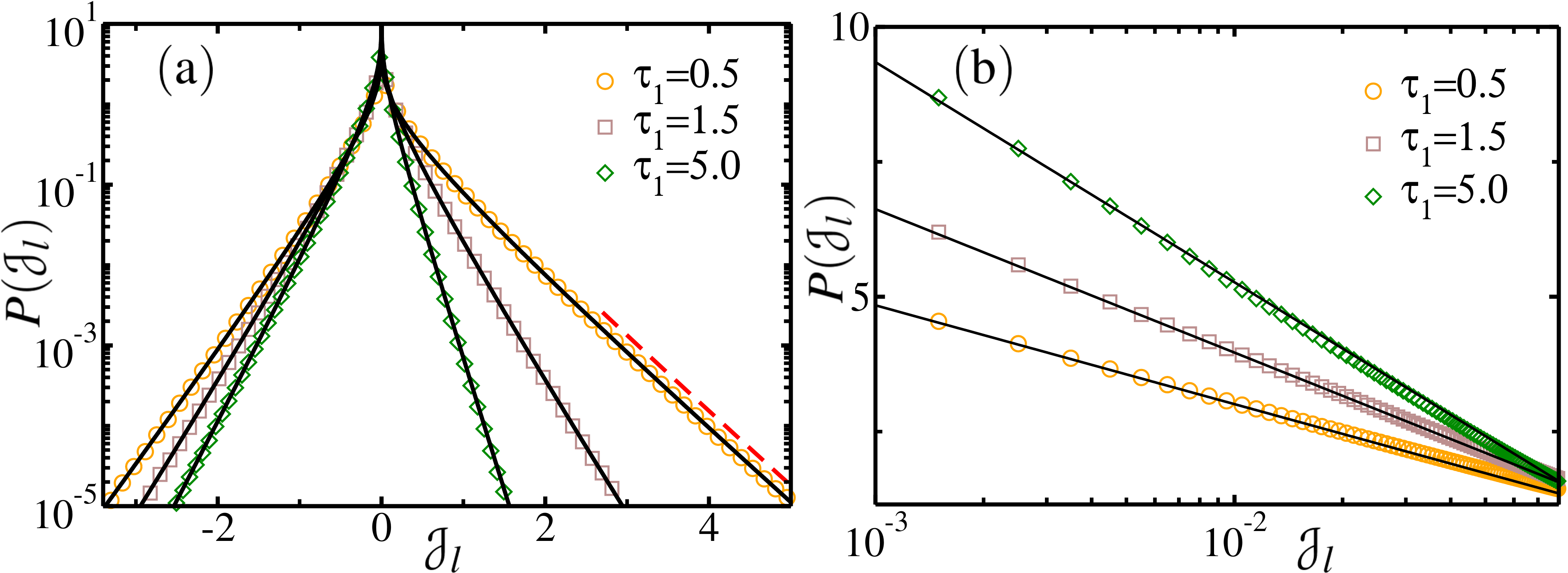}
\caption{(a) Plot of the instantaneous bulk current distribution $P(\mathscr{J}_l)$ for the AOUP driven chain with $\tau_N=1.50$ and different values of $\tau_1$. Black solid lines correspond to the analytical prediction Eq.~\eqref{eq: cur_dist_bulk} and the red dashed line corresponds to the exponential decay predicted in Eq.~\eqref{aoup_jdist_tail}. (b) Plot of $P(\mathscr{J}_l)$ near $\mathscr{J}_l=0$. Black solid lines correspond to Eq.~\eqref{aoup_jdist_origin}.
For both (a) and (b), the symbols correspond to numerical simulations performed on a chain of $N=512$ with  $m=D_1=D_2=\gamma=k=1$.}
\label{jDist_bulk_aoup}
\end{figure}

Using Eq.~\eqref{eq: joint_dist_bulk} and \eqref{eq: moment_gen_bulk} and performing the Gaussian integrals, we get,
\begin{eqnarray}
\langle e^{i \mu \mathscr{J}_l}\rangle &=&\Big[\frac{1}{g_l}\Big(\mu -i\frac{u_l+J_{\text{act}}}{g_l}\Big)\Big(\mu +i\frac{u_l-J_{\text{act}}}{g_l}\Big)  \Big]^{-\frac{1}{2}},\label{eq: ft}
\end{eqnarray}
where $g_l$ and $u_l$ denote stationary correlations, defined as,
\bea
u_{l}=\frac{k}{2}\left[\left\langle \left(v_{l-1}+v_{l}\right)^2 \right\rangle \Big \langle \left(x_{l-1}-x_{l}\right)^2 \Big \rangle\right]^{\frac{1}{2}},~
g_{l}= u_l^2-J^2_\text{act}.~~\label{eq:glul_prim}
\end{eqnarray}
Here $J_\text{act}$ is the average active energy current given in Eq.~\eqref{eq:J}.

The current distribution can be exactly computed by taking the inverse Fourier transform of Eq.~\eqref{eq: ft} [see Appendix~\ref{appendix_curdist} for the detail] which yields,
\begin{equation}
P(\mathscr{J}_l) = \frac{1}{\sqrt{\pi^2 g_{l}}}e^{\frac{J_\text{act}}{g_{l}} \mathscr{J}_l} K_0\left(\frac{u_{l}}{g_{l}}|\mathscr{J}_l|\right), \label{eq: cur_dist_bulk}
\end{equation}
where $K_0(z)$ is the zeroth order modified Bessel function of second kind.

\begin{figure}[t]
\includegraphics[width=8.8cm]{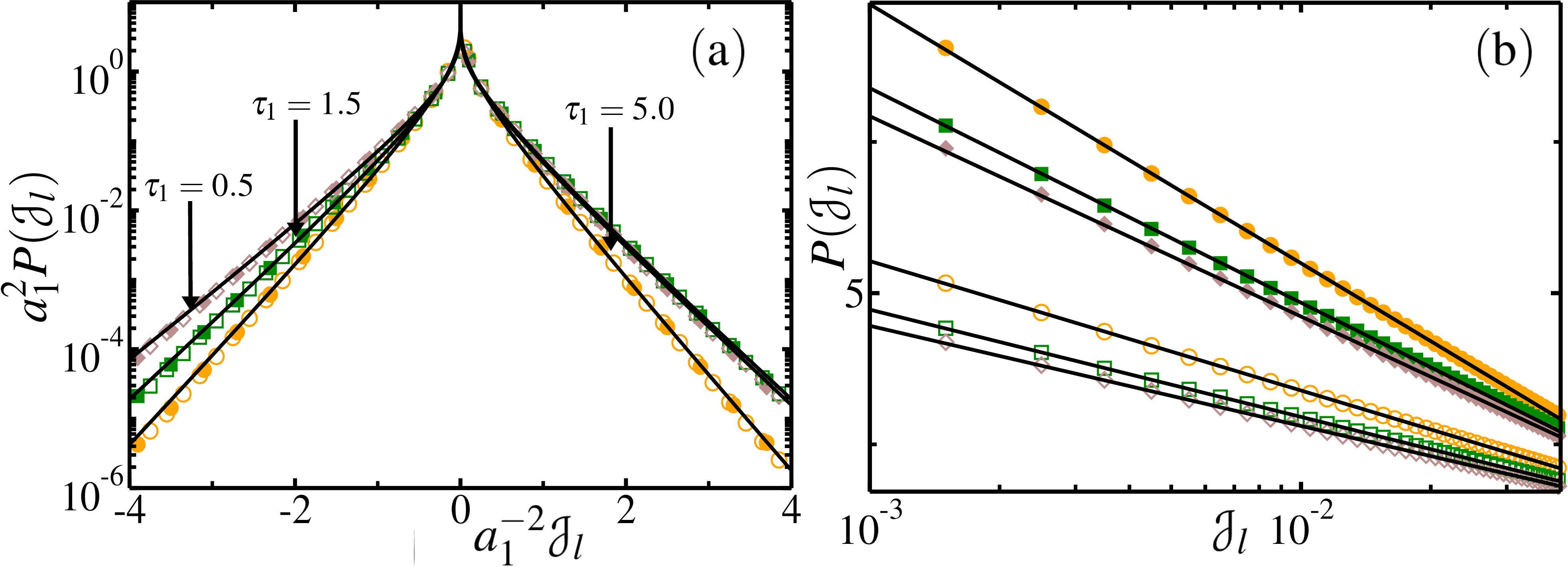}
\caption{(a) Plot of the distribution of the scaled bulk instantaneous current $\mathscr{J}_l/a_1^2$ for RTP (open symbols), and ABP (filled symbols) driven chains for $\tau_N=1.50$ and different values of $\tau_1$. Black solid lines correspond to  Eq.~\eqref{eq: cur_dist_bulk}. (b) Plot of the same data as in (a) zoomed near the origin. The black solid lines here correspond to the logarithmic behavior predicted by Eq.~\eqref{aoup_jdist_origin}. The simulations are performed on a chain of $N=512$ oscillators  with $A_1=A_N=1$ and $D_1^R=D^R_N=1$ for ABP; $m=\gamma=1=k$ here.}
\label{fig:scaled}
\end{figure}

In the thermodynamic limit, $g_l$ can be computed explicitly [see Appendix~\ref{appendix_curdist}] and is given by,
\bea
g_l=\frac{k}{2}\Big( 1+\frac{1}{m}\Big)\hat{T}_{\text{bulk}}^2+\frac{\hat{T}_{\text{bulk}}}{8 \gamma}\left(\frac{a_1^2 \eta_1}{\tau_1}+\frac{a_N^2 \eta_N}{\tau_N}\right)-J_{\text{act}}^2,\label{g_thermo}
\eea
where $\eta_j=\left(1+\frac{4 k \tau_j^2}{m}\right)^{-1/2}-1$ and $\hat{T}_\text{bulk}$ is given in Eq.~\eqref{eq:bulk_kin_temp}. 

Fig.~\ref{jDist_bulk_aoup}(a) compares the numerically measured $P(\mathscr{J}_l)$ at $l=N/2$  with the analytical prediction Eq.~\eqref{eq: cur_dist_bulk} and shows excellent agreement. Interestingly, current distribution is asymmetric and shows divergence near $\mathscr{J}_l=0$, despite having a nonzero mean. In fact, from Eq.~\eqref{eq: cur_dist_bulk}, using the asymptotic behavior of $K_0(z)$ for $z\to 0$, we get,
\bea
P(\mathscr{J}_l)=-\frac{1}{\sqrt{\pi^2 g_l}}\left(\ln{\frac{u_l}{2 g_l}|\mathscr{J}_l|}+E_{\gamma} \right)+O(\mathscr{J}_l)\label{aoup_jdist_origin}
\eea
near $\mathscr{J}_l=0$. Here $E_{\gamma} \simeq 0.577216 $ is the Euler's constant. This logarithmic divergence is illustrated in Fig.~\ref{jDist_bulk_aoup}(b) for different values of the activity. On the other hand the $P(\mathscr{J}_l) $ shows asymmetric exponential decay at the tails.
\begin{figure*}[th]
\includegraphics[width=17cm]{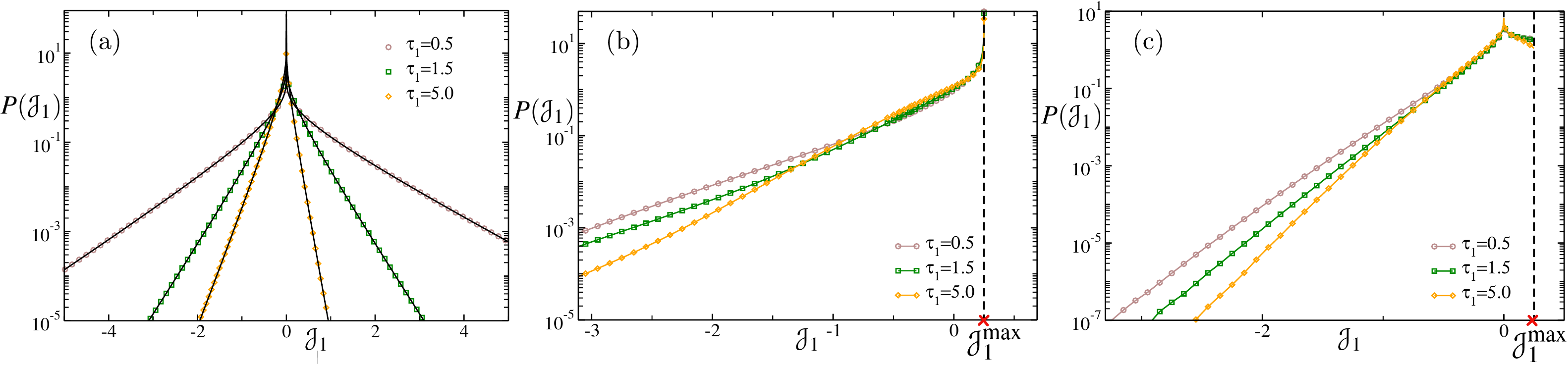}
\caption{Plot of the instantaneous boundary current distribution $P(\mathscr{J}_1)$ for $\tau_N=1.50$ and different values of $\tau_1$ for (a) AOUP, (b) RTP and (c) ABP driven chains. The symbols correspond to the data obtained from numerical simulations performed on a chain of $N=512$ oscillators with $D_1=D_N=1$ for AOUP, $A_1=A_N=1$ for RTP, and $D_1^R=D_N^R$ for ABP. The other parameters are $\gamma=k=m=1$. In (a), the black solid line corresponds Eq.~\eqref{eq:jdist_bound1}. In (b) and (c) the dashed lines indicate the upper bound $\mathscr{J}_1^\text{max}$, see Eq.~\eqref{j1max}.}
\label{fig:jdist_bound}
\end{figure*}

\bea
P(\mathscr{J}_l)\simeq \left\{ 
\begin{split}
&\frac{1}{\sqrt{2 \pi u_l \mathscr{J}_l}}\exp{\left[\frac{J_\text{act}\,\mathscr{J}_l-u_l\mathscr{J}_l}{g_l}\right]}\quad\text{for } \mathscr{J}_l\gg 1,
\cr
&\frac{1}{\sqrt{-2 \pi u_l \mathscr{J}_l}}\exp{\left[\frac{J_\text{act}\,\mathscr{J}_l+u_l\mathscr{J}_l}{g_l}\right]}\quad\text{for } \mathscr{J}_l\ll -1.
\end{split}
\right.
\label{aoup_jdist_tail}
\eea

It should be mentioned here, that form of the distribution \eqref{eq: cur_dist_bulk} is the same as the ones obtained previously in the context of time-integrated heat current fluctuations of Brownian particles in an active environment \cite{heat_fluct2} and relaxation of harmonic oscillators subjected to a temperature quench \cite{heat_fluct1}.

For RTP and ABP driven chains, the current distribution cannot be computed exactly since $\mathscr{P}(x_l,v_l)$ is not known explicitly. However, as we have shown in Sec.~\ref{sec:veldist}, the velocity distribution of the bulk oscillators $P(v_l)$ is Gaussian even for these cases, and one can then expect Eq.~\eqref{eq: joint_dist_bulk} to hold approximately for $1 \ll l \ll N$. In that case, the bulk current distribution for ABP and RTP driven chains should also follow Eq.~\eqref{eq: cur_dist_bulk}. We investigate the validity of this approximation using numerical simulations --- Fig.~\ref{fig:scaled}(a) and (b) compare the measured instantaneous current distribution for ABP and RTP drivings with Eq.~\eqref{eq: cur_dist_bulk}. Indeed, a very good agreement is observed, including the logarithmic divergence near the origin, validating our analytical prediction, for all the three different active drivings. 

The higher moments of the bulk current can, in principle, be calculated from Eq.~\eqref{eq: cur_dist_bulk}. In particular, the second moment is given by [see Appendix~\ref{appendix_j2}],
\bea
\la \mathscr{J}_l^2 \ra&=& 2 J^2_{\text{act}}+u_{l}^2.
\eea
We compare this prediction with numerical simulations in Fig.~\ref{fig:scaled}, which again show a very good agreement, even for ABP and RTP driven chains.

\begin{figure*}[th]
\includegraphics[width=17cm]{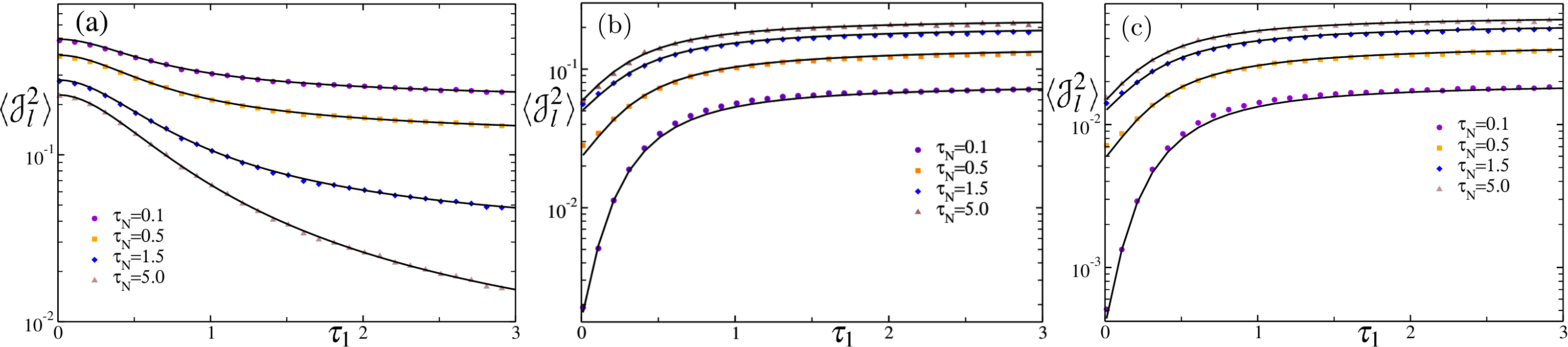}
\caption{Plot of the second moment of the bulk energy current $\mathscr{J}_l^2$ as functions of the activity $\tau_1$, for (a) AOUP, (b) RTP, and (c) ABP driven chains, for different values of $\tau_N$. The symbols correspond to the data obtained from numerical simulations performed on a chain of $N=512$ oscillators with $D_1=D_N=1$ for AOUP, $A_1=A_N=1$ for RTP, $D_1^R=D_N^R=1$ for ABP and $\gamma=k=m=1$. Black solid lines corresponds to the analytical prediction Eq.~\eqref{J2_bulk}.}
\label{fig:j2_bulk}
\end{figure*}

\subsection{Instantaneous current distribution at boundary}
\label{sec:boundary_dist}
The signatures of activity become apparent in the current fluctuation near the boundary. Using the definition of the boundary current $\mathscr{J}_{1}$ given in Eq.~\eqref{boundary_current}, the corresponding  stationary distribution can be written as, 
\bea
P(\mathscr{J}_1)&=&\int dv_1 df_1 \, \delta [\mathscr{J}_1-(-\gamma v_1+f_1) v_1]\,\mathscr{P}(v_1,f_1). \label{eq: j_dist_bound}
\eea
For AOUP driven chain, we can again use the Gaussian nature of the driving force to write,
\begin{eqnarray}
\mathscr{P}(v_1,f_1)=\frac{\exp\left[ -\frac{1}{2}W_1^T\Sigma^{-1}_1 W_1  \right]}{\sqrt{(2 \pi)^2 \text{det}(\Sigma_{1})}},\label{eq: joint_dist_bound}
\end{eqnarray}
where $W_1^T=(v_1~~f_1)$ and $\Sigma_1$ is the corresponding correlation matrix [see Appendix~\ref{ap:boundary_J}]. To obtain $P(\mathscr{J}_1)$, we proceed in the same manner as in Sec.~\ref{sec:bulk_dist} and first compute the moment generating function,
\bea
\langle e^{i \mu  \mathscr{J}_1}\rangle &=& \int dv_{1} df_{1} e^{i \mu (-\gamma v_{1}+ f_{1})v_{1}} \mathscr{P}(v_{1}, f_{1}). 
\eea
Performing the Gaussian integrals, we arrive at an expression which is very similar to the moment generating function of the bulk current, 
\bea 
\langle e^{i \mu  \mathscr{J}_1}\rangle &=&\Big[\frac{1}{g_1}\Big(\mu -i\frac{u_1+J_\text{act}}{g_1}\Big)\Big(\mu +i\frac{u_1-J_\text{act}}{g_1}\Big)  \Big]^{-\frac{1}{2}}
\eea
where,
\begin{eqnarray}
u_1 = \left[\Big( \frac{D_1}{\tau_1}-2 \gamma J_\text{act}  -\gamma^2\hat{T}_1  \Big)\hat{T}_1 \right]^{\frac{1}{2}},~ g_{1}= u_1^2-J^2_\text{act},
\end{eqnarray}

Once again, we can compute the inverse Fourier transform exactly [see detail Appendix~\ref{appendix_curdist}] which yields an explicit form for the boundary current distribution,
\begin{eqnarray}
P(\mathscr{J}_1) = \frac{1}{\sqrt{\pi^2 g_{1}}}e^{\frac{J_\text{act}}{g_{1}} \mathscr{J}_1} K_0\left(\frac{u_{1}}{g_{1}}|\mathscr{J}_1|\right).\label{eq:jdist_bound1}
\end{eqnarray}
$P(\mathscr{J}_{N+1})$ can be computed exactly following the same procedure. Clearly, the shape of boundary current distribution is qualitatively similar to that at the bulk for the AOUP driven chain. In Fig.~\ref{fig:jdist_bound}(a), numerically measured $P(\mathscr{J}_1)$ is plotted along with the analytic curve  Eq.~\eqref{eq:jdist_bound1}, which, as expected, shows an excellent agreement. 

For RTP and ABP driven chains, however,  the distributions of boundary currents  are drastically different. Figure~\ref{fig:jdist_bound}(b) shows $P(\mathscr{J}_1)$ for RTP driven chain which has a monotonically increasing shape and reaches a maximum at $\mathscr{J}_1=\mathscr{J}^{max}_1$, which is independent of $\tau_1$ and $\tau_N$. It also appears that, $P(\mathscr{J}_1$ has a semi-finite support -- it vanishes for $\mathscr{J}_1>\mathscr{J}^{max}_1$. For ABP, on the other hand, the distribution shows a maximum at $\mathscr{J}_1=0$ although the finite cutoff at $\mathscr{J}_1=\mathscr{J}^{max}_1$ is still present in this case. It is hard to compute $P(\mathscr{J}_1)$ in these two cases. However, the existence of the finite cutoff directly follows from the boundedness of the active force  $f_j$ for RTP and ABP. In fact, from the definition of $ \mathscr{J}_{1}=(-\gamma v_{1} + f_{1})v_{1}$, it is clear that $ \mathscr{J}_{1}$ reaches its maximum value for $v_1=f_1^{\text{max}}/2 \gamma$ where $f_1^{\text{max}}$ denotes the maximum value of the active force. This in turn, leads to 
\bea
\mathscr{J}_{1}^{max}= \frac{(f_1^{\text{max}})^2}{4 \gamma}=\frac{A_1^2}{4 \gamma}.\label{j1max}
\eea
This upper cutoff is indicated in Fig.~\ref{fig:jdist_bound}(b) and (c) with  vertical dashed lines, which perfectly agree with the numerically measured distributions. Using a similar argument, one can show that the instantaneous current at the right boundary has a lower cut-off at $\mathscr{J}_{N+1}^{min}$.

\section{Conclusions}\label{sec:concl}
In this work, we study the stationary state properties of a harmonic chain driven by active reservoirs, which exert exponentially correlated stochastic force on the boundary oscillators. Considering three different dynamics of the active force, namely the active Ornstein-Uhlenbeck process, Run-and-Tumble process and active Brownian process, we show that the typical stationary state behavior of the bulk oscillators does not depend on the specific driving. In fact, the bulk kinetic temperature, potential temperature, local velocity and instantaneous current distributions which we compute analytically, all show the same qualitative features irrespective of the specific form of activity driving. Surprisingly, in spite of the inherently nonequilibrium nature of the driving, the velocity distribution of the oscillators at the bulk is Gaussian for all the three different drivings. The shape of the bulk current distributions also turns out to be universal, with a logarithmic divergence near the origin and asymmetric exponential tails. Moreover, the bulk kinetic temperature turns out to be the same as the bulk potential temperature which indicates an equipartition of energy in the bulk of the system. On the other hand, the behavior of the oscillators near the boundaries bear clear signatures of the specific active driving. In fact, unlike the bulk current, the current at the boundary turns out to have a semi-finite bound for RTP and ABP driven chains, which we also compute analytically.    

This work adds a significant step towards the understanding of the activity driven transport. It would be interesting to study the dynamical behavior of the activity driven chain, in particular, the relaxation to the stationary state and how it differs from the thermally driven scenario. Another relevant question is, how does the NESS change when the active reservoirs have more than one time-scale \cite{drabp,drabp_2}. It is also worthwhile to ask how the stationary state behavior changes if the reservoirs are modeled by an extended active particle chain similar to \cite{gupta_21,prashant_activechain}.

\begin{acknowledgments}
The authors would like to thank Abhishek Dhar for useful discussions. R.S. acknowledges support from the Council of Scientific and Industrial Research, India [Grant No. 09/0575(11358)/2021-EMR-I]. U.B. acknowledges support from the
Science and Engineering Research Board (SERB), India,
under a Ramanujan Fellowship [Grant No. SB/S2/RJN-077/2018].
\end{acknowledgments}

\appendix
\section{Matrix formulation and Green's function}
\label{calc_appendix}
The Langevin equations~\eqref{ts} can be solved using a matrix Green's function method \cite{Dhar2001}. For the sake of completeness we provide the detailed solution in this section. It is convenient to recast Eqs.~\eqref{ts} as,
\bea
M\ddot{X}=-\Phi X(t) -\Gamma \dot{X}(t) + F(t), \label{eq:matrix1}
\eea
where $X^T=(x_1~ x_2~\dotsc~x_N )$ is the displacement vector; $M$ and $\Gamma$ are $N$-dimensional matrices with $M_{ij}=m\delta_{ij}$ and $\Gamma_{ij}=\gamma\left(\delta_{i1}\delta_{j1}+\delta_{iN}\delta_{jN}\right)$ and $F(t)$ is an $N$-dimensional column vector with $F_j(t)=f_1(t)\delta_{j1}+f_N(t)\delta_{jN}$; $\Phi$ is a tridiagonal matrix with elements
\begin{eqnarray}
\Phi_{i j}&=&
\begin{cases}
      2 k & \text{for $i=j$},\\
      -k & \text{for $j=i\pm1$ }.
      \end{cases} \label{phi}
\end{eqnarray} 
Equation~\eqref{eq:matrix1} can be solved exactly using the Fourier transform,
\bea
\tilde{X}(\omega)=\int^{\infty}_{-\infty}dt e^{i\omega t}X(t),~\text{and}~ X(t)=\int^{\infty}_{-\infty}\frac{d\omega}{2 \pi} e^{-i\omega t}\tilde{X}(\omega). \qquad\label{ftpos}
\eea
In the frequency domain, Eq.~\eqref{eq:matrix1} reduces to an algebraic equation,
\bea
\tilde{X}(\omega)=G(\omega) \tilde{F}(\omega),\label{sol:FT}
\eea
where $G(\omega)$ is the Green's function matrix defined by,
\bea
G(\omega)=[-M\omega^2+\Phi-i\omega\Gamma]^{-1}.\label{form_G}
\eea
and $\tilde{F}(\omega)$ is the Fourier transform of the active force vector $F(t)$.  The exponential auto-correlation of $F(t)$ leads to,
\bea
\langle \tilde F(\omega) \tilde F^T(\omega') \rangle_{ij}
&=& 2\pi \delta(\omega + \omega')\Big[ \tilde{g}(\omega,\tau_1)\delta_{i1}\delta_{j1}\cr
 && \quad +\tilde{g}(\omega,\tau_N)\delta_{iN}\delta_{jN}\Big], \label{colorcorr}
\eea
where,
\bea
\tilde{g}(\omega,\tau_j) = \frac{2 a_j^2 \tau_j }{1+\omega^2 \tau_j^2}. \label{eq:g_auto}
\eea
From Eq.~\eqref{form_G}, it is clear that $G(\omega)$ is a symmetric matrix and its complex conjugate $G^*(\omega)=G(-\omega)$. The elements of $G$ can be obtained exploiting the tridiagonal structure of $G^{-1}$~\cite{tridiagonal}. In particular, we will need, 
\bea
G_{l1}=k^{l-1}\frac{\theta_{N-l}}{\theta_N},\quad\text{and}\quad
G_{lN}=k^{N-l}\frac{\theta_{l-1}}{\theta_N},\label{G_theta_rel}
\eea
where $\theta_l$ satisfies recursion relations,
\bea
\theta_l &=& (-m \omega^2 +2k)\theta_{l-1}-k^2\theta_{l-2},\quad l = 2,3,\cdots N-1, \quad \\
\theta_{N}&=&(-m\omega^2+2k-i\omega \gamma)\theta_{N-1}-k^2\theta_{N-2},\label{recur1}
\eea
with the boundary conditions, $\theta_0 = 1$ and $\theta_1 = -m\omega^2 + 2k-i\omega\gamma$. The above recursion relations can be explicitly solved to get,
\bea
\theta_l &=& \frac{(-k)^{l-1}}{\sin{q}}[k \sin{(l+1)q}-i\omega \gamma \sin{lq}],~ \text{for}~~ 2 \le l \le N-1, \cr
&& \label{recur2} \\
\theta_N &=& \frac{(-k)^N}{\sin{q}}[a(q)\sin{Nq}+b(q)\cos{Nq}], \label{theta_N}
\eea
where $\omega$ and $q$ are related through,
\bea
\cos{q}=\left(1-\frac{m\omega^2}{2k}\right),\quad \text{and}~ \omega= \omega_c \sin{\frac q2}, \label{eq:omega}
\eea
with $\omega_c=2\sqrt{k/m}$.
Moreover, for notational simplicity, we have introduced,
\bea
a(q) &=& -\frac{2i\gamma \omega}{k}+\cos{q}\left(1-\frac{\gamma^2 \omega^2}{k^2} \right),\n\\
\text{and }
b(q) &=& \sin{q}\left( 1+\frac{\gamma^2 \omega^2}{k^2}\right).\label{eq:ab}
\eea
Note that, for $|\omega|<\omega_c$ i.e. for frequencies within the characteristic band of the harmonic chain, $q\in[-\pi,\pi]$, whereas for $|\omega|>\omega_c$, $q$ becomes complex.

\section{Velocity correlations} \label{appendix_vel_cross_corr}
In this section, we provide the details of the computation of the nearest neighbor velocity correlations $\la v_{l-1}(t) v_l(t) \ra=\la \dot{X}(t) \dot{X}^T(t) \ra_{l-1,l}$ in the steady state. To this end, using \eqref{ftpos} and \eqref{sol:FT}, we first note that,
\bea
\dot{X}(t)=\int^{\infty}_{-\infty} \frac{d \omega}{2 \pi}(-i \omega) e^{-i \omega t} G(\omega) \tilde{F}(\omega).
\eea 
Using the above equation along with Eq.~\eqref{colorcorr} we get,
\bea
\la v_{l-1}(t)v_l(t)\ra &=& \int_{-\infty}^{\infty} \frac{d \omega}{2 \pi} \omega^2 \Big[ G_{l-1,1}G_{1l}^* \tilde{g}(\omega,\tau_1)\n\\
&&~~~+ G_{l-1,N}G_{Nl}^* \tilde{g}(\omega,\tau_N) \Big].
\eea
Thus it is clear that there are two separate contributions from the left and right reservoirs. In the following we explicitly compute the contribution coming from the left reservoir, 
\bea
V_l(\tau_1)=\int \frac{d \omega}{2 \pi} \omega^2 G_{l-1,1}G_{1,l}^*\tilde{g}(\omega,\tau_1),\label{vtau1}
\eea
and the contribution from the right reservoir can be computed similarly. Using Eq.~\eqref{G_theta_rel}, we have,
\bea
V_l(\tau_1)=(-k)^{2l-3} \int_{-\infty}^{\infty}\frac{d \omega }{2 \pi} \omega^2 \frac{\theta_{N-l+1} \theta^*_{N-l}}{|\theta_N|^2}\tilde{g}(\omega,\tau_1).
\eea
Clearly, $V_l(\tau_1)$ will have non-zero contributions from only the even components of the integrand. Hence, using explicit forms of $\theta_l$ and $\theta_n$ from Eqs.~\eqref{recur2} and \eqref{theta_N} and keeping only the terms which are even in $\omega$, we get, 
\bea
V_l(\tau_1)&=&\frac{1}{\pi k^4}\int^{\infty}_0 d \omega\,  \frac{\omega^2\sin(Nq-lq) \sin(Nq-lq+q)}{|a(q)\sin{N q}+b(q)\cos{N q}|^2}\n \\ &&~~~~~~~~~~~~~\times (k^2+\omega^2 \gamma^2) \tilde{g}(\omega,\tau_1).
\eea
Finally, since we are interested in calculating the correlation function in the bulk, we take $l=N/2+\epsilon$ and take the limit $\epsilon \ll N$ to get,
\bea
V_l(\tau_1)&=&\frac{1}{2 \pi k^4}\int^{\infty}_0 d \omega \frac{\omega^2\Big[\cos{q}-\cos{\Big((N+2 \epsilon+1)q\Big)}\Big]}{|a(q)\sin{N q}+b(q)\cos{N q}|^2}\n\\ &&~~~~~~~~~~~~~~\times (k^2+\omega^2 \gamma^2) \tilde{g}(\omega,\tau_1).\label{v_cross_corr}
\eea
At this point, it is important to remember that, for $\omega>\omega_c$, $q$ becomes complex. Thus, in the large $N$ limit, the integrand vanishes exponentially as $e^{-2 N \bar{q}}$ in the region $\omega>\omega_c$ (where $\bar{q}$ is real). Therefore, the range of the integration reduces to $0 \leq \omega \leq \omega_c$, or in terms of $q$, $0\leq q \leq \pi$.
Moreover, in the thermodynamic limit, $\sin{Nq}$ and $\cos{Nq}$ are highly oscillatory and the resulting integrand can be well approximated by averaging over the fast oscillations in $x=Nq$~\cite{kannan_12}. This averaging can be performed using the following identities,
\bea
\frac{1}{2 \pi} \int^{2\pi}_0\frac{d x}{(c_1 \sin{x}+d \cos{x})^2+c_2^2 \sin{x}^2}&=&-\frac{1}{c_2 d},~\text{for}~c_2<0,\n\\
\frac{1}{2 \pi} \int^{2\pi}_0\frac{d x\cos{x}}{(c_1 \sin{x}+d \cos{x})^2+c_2^2 \sin{x}^2}&=&0. \label{ap:fastos}
\eea
Identifying $c_1$, $c_2$ and $d$ as the real and imaginary parts of $a(q)$, and real part of $b(q)$, respectively [see Eq.~\eqref{eq:ab}], we get,
\bea
V_l(\tau_1)=\frac{1}{4 \pi k \gamma} \int^{\infty}_0 dq~\omega \left|\frac{d\omega}{dq} \right|\cot{q}~\tilde{g}(\omega,\tau_1).
\eea
The above integral can be performed exactly using the explicit form of $\omega(q)$ and $g(\omega,\tau_1)$ from Eq.~\eqref{eq:omega} and Eq.~\eqref{eq:g_auto}. Similarly, the contribution from the right reservoir $V_l(\tau_N)$ can also be calculated. Combining these results, we finally get, 
\bea
\la v_{l-1} v_l \ra=\frac{\hat{T}_{\text{bulk}}}{m}+\frac{1}{4 k \gamma}\left(\frac{a_1^2 \eta_1}{\tau_1}+\frac{a_N^2 \eta_N}{\tau_N}\right),\label{eq: vel_corr}\eea
where $\eta_j = \left(1+\frac{4 k \tau_j^2}{m}\right)^{-1/2}-1$.

\section{Potential Energy profile}\label{appendix_pot}

In this section, we explicitly compute the average potential energy of the $l$-th oscillator in the NESS, defined by Eq.\eqref{eq:Uav_def}. As mentioned in Eq.~\eqref{eq:bulk_pot_temp_U}, the average potential energy of the $l$-oscillator can be written as $\frac k4 \sum_{j=1,N}\mathscr{U}_j(l,\tau_j)$, where, $\mathscr{U}_j(l,\tau_j)$ denotes the contribution from the $j$-th reservoir. Using Eq.~\eqref{crosscorr_x} in the definition~\eqref{eq:Uav_def}, we have, for $l\ne 1,N$,
\bea
\mathscr{U}_j(l,\tau_j)&=&\int_{-\infty}^{\infty} \frac{d\omega}{2 \pi} \Big[2 |G_{lj}|^2+|G_{l-1,j}|^2+|G_{l+1,j}|^2 \n \\&-&2 G_{lj} G_{j,l-1}^*-2 G_{lj} G_{j,l+1}^*\Big]\tilde{g}(\omega,\tau_1),\label{u1_bulk}
\eea
where $j=1,N$. On the other hand, for the boundary oscillator $l=1$, we have,
\bea
\mathscr{U}_j(1,\tau_j)&=&\int_{-\infty}^{\infty} \frac{d \omega }{2 \pi }  \Big[3 |G_{1j}|^2
+|G_{2j}|^2 -2 G_{1j}  G_{j\,2}^*\Big]\tilde{g}(\omega,\tau_1).\cr
&& \label{u1_boundary}
\eea
Note that the Fourier transform of the two-point autocorrelation function of colored noise $\tilde{{g}}(\omega)$ is an even function of $\o$, therefore in Eqs.~\eqref{u1_bulk} and \eqref{u1_boundary}, we can neglect the terms with odd power of $\o$ as they would give vanishing contribution. In the following,  we compute the non-zero contributions explicitly, for a thermodynamically large chain.

\subsection{Potential energy for the bulk oscillators ($1 \ll l \ll N$)}
We start with the computation of $U_l$ for the oscillators at the bulk, \ie, for $1 \ll l \ll N$. Using Eqs.~\eqref{G_theta_rel}, \eqref{recur2}, and \eqref{theta_N} in Eq.~\eqref{u1_bulk} we have,
\bea
\mathscr{U}_1(l,\tau_1)&=&\frac{1}{\pi k^4}\int_{0}^{\infty} d\omega \frac{I_1(l,\omega)\tilde{g}(\omega,\tau_1)}{|a(q) \sin{Nq}+b(q)\cos{Nq}|^2},~~~~
\eea
where, we have kept only the even component of the integrand in \eqref{u1_bulk}, with 
\bea
I_1(l,\omega) &=&4 \sin^2{\frac{q}{2}} \Big [ \gamma^2 \omega^2 \cos{q}  \cos{(2Nq-2lq)}+ (k^2+\gamma^2 \omega^2) \cr
&& +   k^2 \, \cos{q} \cos{(2Nq-2 lq+2q)}\Big]. \label{Il}
\eea
for $l \ne 1,N$. Similarly, the contribution from the right reservoir can be expressed as,
\bea
\mathscr{U}_N(l,\tau_N)&=&\frac{1}{\pi k^4}\int_{0}^{\infty} d\omega \frac{I_N(l,\omega)\tilde{g}(\omega,\tau_N)}{|a(q) \sin{Nq}+b(q)\cos{Nq}|^2}\n\\
\text{where, }I_N(l,\omega)&=&I_1(N-l+1,\omega).
\eea
It is easy to see that, in the thermodynamic limit, the integrand vanishes in the region $\omega>\omega_c$ [see the discussion after Eq.~\eqref{v_cross_corr}]. Moreover, averaging over the fast oscillations in this limit using the identities \eqref{ap:fastos}, we get, 
\bea
\mathscr{U}_1(l,\tau_1)=\frac{1}{\pi k\gamma}\int^{\pi}_0 dq \left|\frac{d \omega}{dq} \right|\frac{(1-\cos{q})}{\omega \sin q} \tilde{g}(\omega,\tau_1),\n 
\eea
Using the $\omega-q$ relation \eqref{eq:omega} and the explicit form of $\tilde g(\omega,\tau)$ from Eq.~\eqref{eq:g_auto}, we arrive at,
\bea
\mathscr{U}_1(l,\tau_1)=\frac{m  \tau_1 a_1^2}{k \gamma \pi}\int^{\pi}_0  \frac{dq}{m+2 k \tau_1^2(1-\cos{q})}.
\eea
This integral can be computed exactly and yields,
\bea
\mathscr{U}_1(l,\tau_1)=\frac{a_1^2 \tau_1}{k\gamma\sqrt{1+\frac{4 k \tau_1^2}{m}}}.
\eea
The contribution from the right reservoir can be similarly obtained and turns out to be of the same form; the final expression of the average potential energy for the oscillators at the bulk is then given by,
\bea
{U}_l=\frac{a_1^2 \tau_1}{4 \gamma\sqrt{1+\frac{4 k \tau_1^2}{m}}}+\frac{a_2^2 \tau_2}{4 \gamma\sqrt{1+\frac{4 k \tau_2^2}{m}}},
\eea
in the thermodynamic limit.

\subsection{Potential energy for the oscillator near the left boundary}

The average potential energy of the left boundary oscillator $U_1$ has two contributions $\mathscr{U}_1(1,\tau_1)$ and $\mathscr{U}_N(1,\tau_N)$ from the reservoirs at the two ends, given by Eq.~\eqref{u1_boundary}. Substituting, the explicit forms of $G_{lm}$ from Eq.~\eqref{G_theta_rel} and \eqref{theta_N}, we get the contribution from the left reservoir,
\bea
\mathscr{U}_1(1,\tau_1)&=&\frac{1}{\pi k^4 }\int^{\infty}_{0} d\omega \frac{ I_1(1,\omega) \tilde{g}(\omega,\tau_1)}{|a(q) \sin{Nq}+b(q)\cos{Nq}|^2}.\qquad \label{bound_1_l}
\eea
with,
\bea
I_1(1,\omega)&=&\gamma^2 \omega^2 \Big[(1-\cos{q}) \cos{(2 Nq-3q)}-\cos{(2Nq-2q)}\Big]\cr
&+& k^2\Big[(1-\cos{q}) \cos{(2 Nq-q)}-\cos{\big(2Nq}\big)\Big]\cr
&+&(k^2+\gamma^2 \omega^2)(2-\cos{q}),\label{I1_outband}
\eea
where, as before, we have kept only the terms which are even in $\omega.$ Similarly, we have the contribution from the right reservoir,
\bea
\mathscr{U}_N(1,\tau_N)&=&\frac{1}{\pi k^4 }\int^{\infty}_{0} d\omega\frac{I_N(1,\omega) \tilde{g}(\omega,\tau_N)}{|a(q) \sin{Nq}+b(q)\cos{Nq}|^2},\qquad~ 
\label{bound_N_l}
\eea
where,
\bea
I_N(1,\omega)&=&(k^2+\gamma^2 \omega^2)(2-\cos{q})+k^2[(1-\cos{q}) \cos{3q}\n\\&-&\cos{2q}]+\gamma^2 \omega^2 [(1-\cos{q}) \cos{q}-1].
\eea
Once again, in the thermodynamic limit $N \to \infty$, the integrand vanishes for $\omega>\omega_c$ and shows fast oscillations for $\omega <\omega_c$. Averaging over these fast oscillations as before, we get, 
\bea
\mathscr{U}_N(1,\tau_N)&=&\frac{1}{8 \pi \gamma k}\int^{\pi}_0 dq \csc^2(q/2)\Big[
 (2-\cos{q})\n\\&+& \frac{k^2[(1-\cos{q})\cos{3q-\cos{2q}}]}{(k^2+\gamma^2 \omega^2)}\n\\&+&\frac{\gamma^2 \omega^2 [(1-\cos{q})\cos{q}-1]}{(k^2+\gamma^2 \omega^2)}\Big]\tilde{g}(\omega,\tau_N).\qquad
\label{num_pot1}
\eea
This integral can be evaluated numerically remembering $\omega=\omega_c \sin(q/2)$ and using $\tilde{g}(\omega,\tau_j)$ from Eq.~\eqref{eq:g_auto}.

In contrast, the contribution from the left reservoir is non-zero for the whole domain $0<\omega<\infty$. In this case, it is convenient to consider the contributions from inside the band ( $0\leq \omega \leq \omega_c$) and outside the band ($\omega>\omega_c$) separately and write,
\bea
\mathscr{U}_1(1,\tau_1)=\mathscr{U}^b_1(1,\tau_1)+\mathscr{U}^o_1(1,\tau_1). \label{ap:sepU}
\eea
The  contribution from inside the band is given by, 
\bea
\mathscr{U}^b_1(1,\tau_1)=\frac{1}{\pi k^4}\int^{\omega_c}_0 d\omega \frac{I_1(1,\omega) \, \tilde{g}(\omega,\tau_1)}{|a(q) \sin{Nq}+b(q)\cos{Nq}|^2}.\qquad
\eea
As before, in the thermodynamic limit, we can average over the fast oscillations in $x=Nq$ to get,
\bea
\mathscr{U}^b_1(1,\tau_1) &=& \frac{1}{\pi k^4 }\int^{\pi}_0 dq \left|\frac{d \omega}{dq}\right|  \Big[\gamma^2 \omega^2[(1-\cos{q})Q(3q)-Q(2  q)]\cr
&+&  k^2[(1-\cos{q}) Q(q) -Q(0)] +\frac{k^3(2-\cos{q})}{2 \gamma \omega \sin{q}}\Big]\tilde{g}(\omega,\tau_1)\cr
&& \label{num_pot2}
\eea
where,
\bea
Q(\nu)&=&\int^{2 \pi}_0\frac{d x}{2 \pi}\frac{\cos(2 x-\nu)}{(c_1 \sin{x}+d\cos{x})^2+c_2^2 \sin{x}^2}\n \\ &=&\frac{\cos{\nu} \left(c_1^2+c_2^2-d^2\right)-2 c_1 d \sin {\nu }}{-c_2 d \left(c_1^2+(c_2-d)^2\right)},
\eea
with $c_2=\text{Im}[a(q)]=-\frac{2\gamma \omega}{k}$ and $d=\text{Re}[b(q)]=\sin{q}\left( 1+\frac{\gamma^2 \omega^2}{k^2}\right)$.

Outside the band, \ie, for $\omega>\omega_c$, $q$ becomes complex, and to compute the corresponding contribution,
\bea
\mathscr{U}^o_1(1,\tau_1)=\frac{1}{\pi k^4 }\int^{\infty}_{\omega_c}d \omega \frac{ I_1(1,\omega) \tilde{g}(\omega,\tau_1)}{|a(q) \sin{Nq}+b(q)\cos{Nq}|^2},\qquad\label{out_band}
\eea
it is convenient to define $q=\pi-i \bar{q}$ where $\bar{q}$ is a real variable and $\omega=\omega_c \cosh{\bar{q}/2}$. The integral in Eq.~\eqref{out_band} takes a much simpler form in terms of $\bar{q}$, 
\bea
\mathscr{U}_{1}^o(1,\tau_1)&=&\frac{\omega_c}{\pi k^4}\int^{\infty}_0d\bar{q}  
\frac{\sinh{\frac{\bar{q}}{2}} \cosh^2{\frac{\bar{q}}{2}} e^{-3\bar{q}}}{ \big(\gamma^2 \omega^2 e^{-2\bar{q}}+k^2\big)} \tilde{g}(\omega,\tau_1).\qquad
\label{num_pot3}
\eea
where we have used Eq.~\eqref{I1_outband} and the identities,
\bea
\sin{nq}=(-1)^{n+1}i \sinh{n\bar{q}},\,\cos{nq}&=&(-1)^{n} \cosh{n\bar{q}},\quad 
\eea
for integer $n$ to average over the fast oscillations in $N\bar{q}$. Finally, the average potential energy of the left boundary oscillator can be evaluated combining the
Eqs.~\eqref{num_pot1}, \eqref{num_pot2} and \eqref{num_pot3}.

For $1< l\ll N/2$ the average potential energy can be computed following a similar procedure and we quote the final results here. In this case, the contribution from the right boundary simplifies to,
\bea
\mathscr{U}_N(l,\tau_N)&=&\frac{1}{2 \pi k \gamma}\int^{\pi}_0 dq \Big[ 1+\frac{\cos{q}}{k^2+\gamma^2 \omega^2}\Big( \gamma^2 \omega^2 \cos(2 l q-2 q)\n\\&&~~~~~~~~~~~~ +k^2 \cos{(2 l q)}\Big) \Big]\tilde{g}(\omega,\tau_N).
\label{num_pot4}
\eea
As before, the left reservoir contribution comprises of two parts---one coming from inside the band,
\bea
\mathscr{U}^b_1(l,\tau_1)&=&\frac{1}{2\pi k^4}\int^{\pi}_0 dq \Big[ \frac{k^3}{ \gamma}+\omega \sin{2 q}\Big( \gamma^2 \omega^2 Q(2 l q) \n \\ &+&  k^2 Q(2lq -2 q) \Big) \Big]\tilde{g}(\omega,\tau_1),
 \label{num_pot5}
\eea
and the other coming from outside the band,
\bea
\mathscr{U}^o_1(l,\tau_1)&=&-\frac{1}{2 \pi k^4}\int^{\infty}_0 d\bar{q} \frac{\omega \sinh{2 \bar{q}\,e^{-2 l \bar{q}}}}{\gamma^2 \omega^2 e^{-2 \bar{q}}+k^2} \tilde{g}(\omega,\tau_1).
\label{num_pot6}\qquad
\eea
which can be computed numerically.

\subsection{$U_l$ near the right boundary}

The average potential energy of the oscillators near the right boundary can be evaluated by exploiting the symmetry of the system. To this end, it is convenient to define $\ell=N-l+1$, where $\ell=1,2,3\cdots \ll N$. In this notation ${U}_{N-\ell+1}$ corresponds to the potential energy profile near the right boundary and can be expressed as,
\bea
{U}_{N-\ell+1}=\frac{k}{4}\Big[\mathscr{U}_N(\ell,\tau_1)+\mathscr{U}_1(\ell,\tau_N)\Big].
\eea
Here $\mathscr{U}_N (\ell,\tau_1)$ and $\mathscr{U}_1 (\ell,\tau_N)$  denotes the contribution from the left and right reservoirs and can be evaluated using Eq.~\eqref{num_pot1}--\eqref{num_pot6}.

\section{Instantaneous current distribution for Active Ornstein Uhlenbeck Process}
\label{appendix_curdist}

In this section we provide the detailed derivation of the instantaneous current distribution for the AOUP driven chain quoted in Eqs.~\eqref{eq: cur_dist_bulk} and \eqref{eq:jdist_bound1}.
\subsection{Distribution of the boundary current}
\label{ap:boundary_J}
Let us start with the left boundary current $\mathscr{J}_1$ which is defined as,
\bea
\mathscr{J}_{1}(t)&=&(-\gamma v_{1}+ f_{1})v_{1}.\label{ap:bondarycur}
\eea
The Gaussian nature of the stationary state for AOUP implies that the joint distribution of $\{v_1,f_1\}$ is a bivariate Gaussian, 
\begin{eqnarray}
\mathscr{P}(v_1,f_1)=\frac{\exp\left[ -\frac{1}{2}W_1^T\Sigma^{-1}_1 W_1  \right]}{\sqrt{(2 \pi)^2 \text{det}(\Sigma_{1})}},\label{a1}
\end{eqnarray}
where $W_1^T=(v_1~~f_1)$ and the correlation matrix,
\begin{eqnarray}
\Sigma_1 &=& \begin{bmatrix}
\langle v^2_{1} \rangle & \langle v_{1} f_{1} \rangle\\
\langle v_{1} f_{1} \rangle &\langle f_{1}^2\rangle\\ 
\end{bmatrix}.
\end{eqnarray}
is positive-definite. Using Eq.~\eqref{ap:bondarycur} the distribution of $\mathscr{J}_1$ can be expressed as,
\bea 
P(\mathscr{J}_1)&=& \int dv_1 df_1 \delta [\mathscr{J}_1-(-\gamma v_1+f_1) v_1]\mathscr{P}(v_1,f_1).
\eea
The corresponding moment generating function, which is nothing but the Fourier transform of $P(\mathscr{J}_1)$, is given by, \begin{eqnarray}
\langle e^{i \mu \mathscr{J}_1}\rangle &=& \int dv_{1} df_{1} \mathscr{P}(v_{1}, f_1) e^{i \mu (-\gamma v_{1}+ f_{1})v_{1} }.\label{generating1}
\end{eqnarray}
where, $\la f_1^2 \ra = D_1/\tau_1$ in the stationary state.

The Gaussian integrals over $v_1$ and $f_1$ can be evaluated exactly using Eq.~\eqref{a1}, yielding, 
\begin{eqnarray}
\langle e^{i \mu \mathscr{J}_1}\rangle &=&\sqrt{ab}\Big[(\mu-ia)(\mu+ib)  \Big]^{-\frac{1}{2}},\n\\
\text{with}~~
a&=&\frac{u_1+J_\text{act}}{g_1}\text{ and }b=\frac{u_1-J_\text{act}}{g_1}.\label{ap:ft}
\end{eqnarray}
Here $u_1$ and $g_1$ correspond to certain stationary state correlations, given by,    
\begin{eqnarray}
u_1 &=& \left[\Big( \frac{D_1}{\tau_1}-2 \gamma J_\text{act}  -\gamma^2\hat{T}_1  \Big)\hat{T}_1 \right]^{\frac{1}{2}},\cr
g_{1}&=&\text{det}(\Sigma_1)= u_1^2-J^2_\text{act}\label{g1_u1}
\end{eqnarray}
where $\hat{T}_1=\la v_1^2 \ra$ is the kinetic temperature of the oscillator at the left boundary which has been calculated in reference \cite{activity_driven_chain} .
\begin{figure}[ht]
      \includegraphics[scale=0.32]{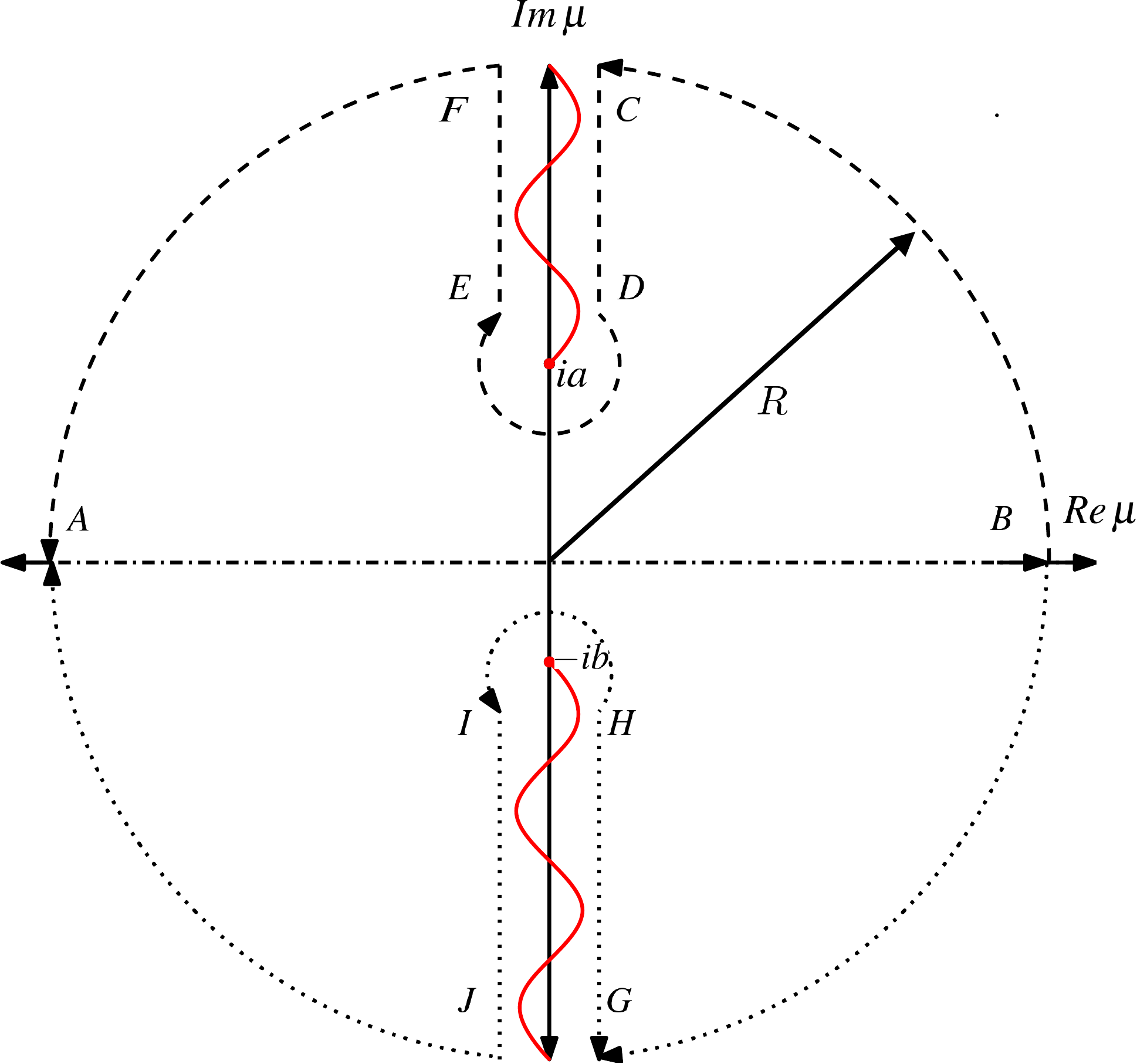}
      \caption{A schematic representation of the contours used to evaluate the complex integral in Eq.~\eqref{contour}. }
      \label{contour_image}
\end{figure}

The current distribution can be obtained by taking the inverse Fourier transform of the moment generating function Eq.~\eqref{ap:ft},
\begin{eqnarray}
P(\mathscr{J}_1)=\sqrt{a\,b}\int _{-\infty}^{\infty}\frac{d \mu}{2 \pi} \frac{e^{-i \mu \mathscr{J}_1} }{\sqrt{(\mu-i a)(\mu+i b)}}.\label{contour}
\end{eqnarray}
To evaluate this complex integral explicitly we need to choose a convenient contour. $a$ and $b$ are real positive quantities {[Eq.~\eqref{g1_u1}]. Hence the integrand in Eq.~\eqref{contour} has two branch points at $\mu=ia$ and $\mu=-ib$. We choose the corresponding branch cuts as shown in Fig.~\ref{contour_image}. Now, for $\mathscr{J}_1<0$ one can draw a closed  contour $ABCDEFA$ which has no singularities inside and hence,
\begin{eqnarray}
I_{A\rightarrow B}+ I_{B\rightarrow C} + I_{C\rightarrow D}+I_{D\rightarrow E}+I_{E\rightarrow F}+I_{F\rightarrow A}=0,\label{complex_int}
\end{eqnarray}
where $I_{\alpha \to \beta}$ denotes the the integral Eq.~\eqref{contour} evaluated along the path $\alpha\to\beta$.

\begin{figure*}[th]
\includegraphics[width=17cm]{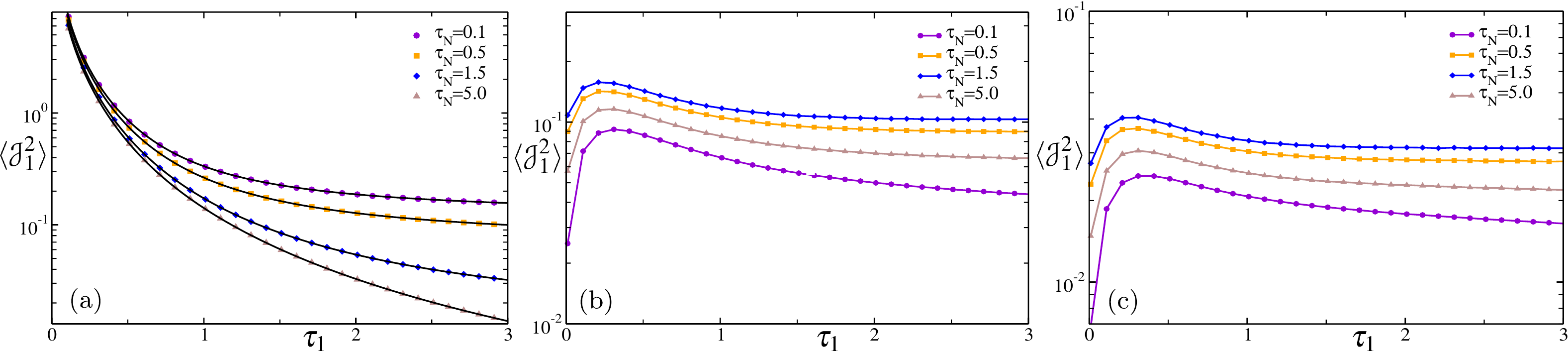}
\caption{Plot of the second moment of the boundary current $\mathscr{J}_1^2$ as functions of the activity $\tau_1$, for (a) AOUP, (b) RTP,  and (c) ABP driven chains, for different values of $\tau_N$. The symbols correspond to the data obtained from numerical simulations performed on a chain of $N=512$ oscillators with $D_1=D_N=1$ for AOUP, $A_1=A_N=1$ for RTP, $D_1^R=D_N^R=1$ for ABP and $\gamma=k=m=1$. Black solid lines in (a) correspond to the analytical prediction Eq.~\eqref{J2_boundary}.}
\label{fig:j2_boundary}
\end{figure*}

Clearly, for $\mathscr{J}_1<0$, the contribution from $I_{B\to C}$ and $I_{F\to A}$ vanish when the radius of the arcs $R\to\infty$. Similarly $I_{D\to E}\to0$ when radius of the circular arc $DE$ vanishes. Hence, from Eq.~\eqref{complex_int} we have,
\bea
P(\mathscr{J}_1)=I_{A\to B}=-\Big[I_{C\to D}+I_{E\to F} \Big].\label{comp1}
\eea   
To evaluate $I_{C\to D}$ and $I_{E\to F}$ we note that along the segments $CD$ and $EF$, $\mu=ia+r e^{\frac{i \pi}{2}}$ and $\mu =ia+r e^{-\frac{i 3\pi}{2}}$, respectively, where $r\in [0,\infty)$. Substituting these in Eq.~\eqref{contour} and using Eq.~\eqref{comp1}, we finally get, for $\mathscr{J}_1<0$,
\bea
P(\mathscr{J}_1)&=& \frac{\sqrt{a\, b}}{\pi} e^{a \mathscr{J}_1} \int^{\infty}_{0} \frac{dr\,e^{\,r \mathscr{J}_1} }{\sqrt{r(r+a+b)}}\n \\ &=& \frac{\sqrt{a\,b}}{\pi} e^{\frac{a-b}{2} \mathscr{J}_1}K_0\left( -\frac{a+b}{2}\mathscr{J}_1\right),\qquad\label{pj1}
\eea
where $K_0(z)$ is the modified Bessel function of second kind. The distribution for $\mathscr{J}_1>0$ can be computed similarly by choosing the contours $ABGHIJA$. In this case, we get, 
\bea
P(\mathscr{J}_1)&=&\frac{\sqrt{a\,b}}{\pi} e^{\frac{a-b}{2} \mathscr{J}_1}K_0\left( \frac{a+b}{2}\mathscr{J}_1\right).\label{pj2}
\eea  
Using the explicit forms of $a$ and $b$ and combining Eqs~\eqref{pj1} and \eqref{pj2} we get the complete boundary current distribution which is quoted in Eq.~\eqref{eq:jdist_bound1}. 

\subsection{Distribution of the bulk current}
The instantaneous current flowing from the $l-1$-th to $l$-th oscillator is defined as,
\bea
\mathscr{J}_{l}&=&\frac{k}{2}\left(v_{l-1}+v_{l}\right) \left(x_{l-1}-x_{l} \right).\label{bulk_cu}
\eea
The distribution $P(\mathscr{J}_l)$ of the instantaneous current at the bulk is then given by,
\bea
P(\mathscr{J}_l)&=&\int dv_{l-1}\,dv_{l}\,dx_{l-1}\,dx_{l}\,  \delta\left[\mathscr{J}_l-\frac{k}{2} ( v_{l-1}+v_{l})( x_{l-1}-x_{l})\right] \n \\ &&\qquad\times \mathscr{P}(v_{l-1}, v_{l}, x_{l-1}, x_{l}).
\eea

Here, $\mathscr{P}(v_{l-1},v_{l},x_{l-1},x_{l})$ denotes the joint position and velocity distribution of the $l-1$-th and $l$-th oscillators which is a multivariate Gaussian, 
\begin{eqnarray}
\mathscr{P}(v_{l-1},v_{l},x_{l-1},x_{l})&=&\frac{\exp\left[ -\frac{1}{2}W_l^T\Sigma^{-1}_l W_l  \right]}{\sqrt{(2 \pi)^4 \text{det}(\Sigma_{l})}},\label{a2}\cr
\text{with, } W_l^T&=&(v_{l-1}~v_{l}~x_{l-1}~x_{l}),
\end{eqnarray}
for the AOUP driven harmonic chain. The correlation matrix $\Sigma_l$ is given by,
\begin{eqnarray}
\Sigma_l &=& \begin{bmatrix}
\langle v^2_{l-1} \rangle & \langle v_{l-1} v_{l} \rangle &  0   & -J_\text{act}/k \\
\langle v_{l-1} v_{l} \rangle &\langle v_{l}^2 \rangle &   J_\text{act}/k  &  0  \\
 0  &  J_\text{act}/k  & \langle x_{l-1}^2 \rangle & \langle x_{l-1} x_{l} \rangle\\
 -J_\text{act}/k  &  0  & \langle x_{l-1} x_{l} \rangle & \langle x_{l}^2 \rangle 
\end{bmatrix}.\quad \label{eq:corrmat_bulk}
\end{eqnarray}
where we have already used the fact that $\langle v_{l} x_{l}\rangle=0 $ and $\langle v_{l-1} x_{l}\rangle=-\langle v_{l} x_{l-1}\rangle=-\frac{ J_\text{act}}{k}$.

We proceed in the same manner as in Sec.~\ref{ap:boundary_J} and compute the moment generating function, 
\bea
\langle e^{i \mu \mathscr{J}_l}\rangle &=& \int dv_{l-1}\,dv_{l} \,dx_{l-1}\,dx_{l} \,e^{i \mu \frac{k}{2} ( v_{l-1}+v_{l})( x_{l-1}-x_{l})}  \n \\ &&\qquad\times  \mathscr{P}(v_{l-1}, v_{l}, x_{l-1}, x_{l}) \cr
&=&\sqrt{a b}\left[(\mu -ia)(\mu +ib)  \right]^{-\frac{1}{2}}\label{ft}
\end{eqnarray}
where
\begin{eqnarray}
a&=&\frac{u_l+J_\text{act}}{g_l},\text{ and }b=\frac{u_l-J_\text{act}}{g_l},\label{ab_bulk}
\end{eqnarray}
with,
\begin{eqnarray}
u_{l}=\frac{k}{2}\left[\left\langle \left(v_{l-1}+v_{l}\right)^2 \right\rangle \Big \langle \left(x_{l-1}-x_{l}\right)^2 \Big \rangle\right]^{\frac{1}{2}},~g_{l}= u_l^2-J^2_\text{act}.\cr \label{ap:gl}
\end{eqnarray}
Clearly, the moment generating function of the bulk current has the same form as that of the boundary current [see \eqref{ap:ft}]. Consequently, the inverse Fourier transform of \eqref{ft} is also of the same form as Eq.~\eqref{pj2}, 
\begin{eqnarray}
P(\mathscr{J}_l) = \frac{\sqrt{a b}}{\pi } e^{\frac{a-b}{2} \mathscr{J}_l} K_0\left(\frac{a+b}{2}|\mathscr{J}_l|\right),
\end{eqnarray}
where $a$ and $b$ are given by Eq.~\eqref{ab_bulk}-\eqref{ap:gl}.

For a thermodynamically large chain, $g_l$ as well as $u_l$ can be evaluated exactly. In this limit, $ k \la (x_{l-1}-x_{l})^2\ra=\hat{T}_\text{bulk}$  and  $\la v_{l-1}^2 \ra=\la v_{l}^2 \ra=\hat{T}_{\text{bulk}}$; see Appendix~\ref{appendix_vel_cross_corr}. Using these results along with Eq.~\eqref{eq: vel_corr} in Eq.~\eqref{ap:gl}, we get a general expression for $g_l$, quoted in Eq.~\eqref{g_thermo},  which is valid irrespective of the specific active dynamics.

\subsection{Second moment of energy current}
\label{appendix_j2}

Higher moments of the active current can, in principle, be computed from Eq.~\eqref{eq: ft} or \eqref{eq: cur_dist_bulk} for the AOUP driven chain. In this case, the second moments of the bulk and boundary currents, respectively, are given by,
\begin{eqnarray}
\la \mathscr{J}_l^2 \ra&=&  - \frac{d^2}{d \mu^2} \la e^{i\mu \mathscr{J}_l} \ra \Big|_{\mu=0}= 2\,J^2_{\text{act}}+u_{l}^2,\label{J2_bulk} \\
\la \mathscr{J}_1^2 \ra&=& - \frac{d^2}{d \mu^2} \la e^{i\mu \mathscr{J}_1} \ra \Big|_{\mu=0} = 2\,J^2_{\text{act}}+u_{1}^2.\label{J2_boundary}
\end{eqnarray}  
Figures~\ref{fig:j2_bulk}(a) and \ref{fig:j2_boundary}(a) show plots of $\la \mathscr{J}_l^2 \ra$ and $\la \mathscr{J}_1^2\ra$ as functions of $\tau_1$, for different values of $\tau_N$ for the AOUP driven chain. 

For RTP and ABP driven chains, as discussed in Sec.~\ref{sec:bulk_dist}, Eq.~\eqref{eq: cur_dist_bulk} describes the fluctuations of the bulk current reasonably well. Hence, we expect Eq.~\eqref{J2_bulk} also to hold in these cases, which indeed is the case, as shown in Fig.~\ref{fig:j2_bulk}(b) and (c).  The boundary current distributions, for RTP and ABP, however, are drastically different [see Sec.~\ref{sec:boundary_dist}] and consequently, Eq.~\eqref{J2_boundary} is not expected to describe the variance of the boundary currents in these scenarios. Hence we take recourse to numerical simulations in this case --- Figure~\ref{fig:j2_boundary}(b) and (c) show plots of numerically measured $\la \mathscr{J}_1^2 \ra$ for the RTP and ABP driven chains, respectively. It turns out that, similar to the behavior of the average current, the second moment also shows non-monotonic behavior in these cases.

\end{document}